\documentclass[lettersize,journal]{IEEEtran}

\usepackage{amsmath,amsfonts}
\usepackage{algorithmic}
\usepackage{algorithm}
\usepackage{array}
\usepackage{textcomp}
\usepackage{stfloats}
\usepackage{url}
\usepackage{verbatim}
\usepackage{graphicx}
\usepackage{cite}
\usepackage{color, colortbl}
\usepackage{booktabs}
\usepackage{amssymb}
\usepackage{multirow}
\usepackage{cleveref}
\usepackage{xcolor}
\usepackage{tikz}
\usepackage[T1]{fontenc} 
\usepackage{amsmath}
\usepackage[cmintegrals]{newtxmath}
\usepackage{bm} 
\usepackage{cuted}

\usepackage[caption=false,font=footnotesize,labelfont=rm,textfont=rm]{subfig}

\usepackage{subfig}
\usepackage{caption}

\usepackage{balance}
\usepackage[accsupp]{axessibility}

\newcommand{\etal}{\emph{et al.}}
\newcommand{\ie}{i.e., }

\newlength\savewidth

\begin{document}

\title{Scale Up Composed Image Retrieval Learning \\ via Modification Text Generation}

\author{
Yinan~Zhou,
Yaxiong~Wang*,
Haokun~Lin,
Chen~Ma,
Li~Zhu*,
Zhedong~Zheng
\thanks{
* indicates corresponding authors.}
\thanks{
Y.~Zhou and L.~Zhu are with the School of Electronics and Information Engineering, Xi'an Jiaotong University, Xi'an 710049, China (e-mail: zyn13572297710@stu.xjtu.edu.cn; zhuli@mail.xjtu.edu.cn).}
\thanks{
Y.~Wang is with the School of Electronics and Information Engineering, Hefei University of Technology, Hefei 230009, China (e-mail: wangyx15@stu.xjtu.edu.cn).
}
\thanks{
H.~Lin is with the School of Artificial Intelligence, University of the Chinese Academy of Sciences, Beijing 101408, China (e-mail: haokun.lin@cripac.ia.ac.cn).
}
\thanks{
C.~Ma, H.~Lin and Y.~Zhou are with the Department of Computer Science, City University of Hong Kong, Hong Kong 999077, China (e-mail: chenma@cityu.edu.hk).
}
\thanks{
Z.~Zheng is with Faculty of Science and Technology, and Institute of Collaborative Innovation, University of Macau, Macau 999078, China (e-mail: zhedongzheng@um.edu.mo).
}
\thanks{
The paper is supported by the Early Career Scheme (No. CityU 21219323) and the General Research Fund (No. CityU 11220324) of the University Grants Committee (UGC), the NSFC Young Scientists Fund (No. 9240127), National Key Research and Development Program of China (2023YFC3321600), the NSFC project under grant No. 62302140, the Fundamental Research Funds for the Central Universities (Academic Newcomer Support Program of Hefei University of Technology with project No. JZ2024HGTB0261), and University of Macau Start-up Research Grant SRG2024-00002-FST and Multi-Year Research Grant MYRG-GRG2024-00077-FST-UMDF.}
}

\maketitle

\begin{abstract}
Composed Image Retrieval (CIR) aims to search an image of interest using a combination of a reference image and modification text as the query. Despite recent advancements, this task remains challenging due to limited training data and laborious triplet annotation processes. To address this issue, this paper proposes to synthesize the training triplets to augment the training resource for the CIR problem.  Specifically, we commence by training a modification text generator exploiting large-scale multimodal models and scale up the CIR learning throughout both the pretraining and fine-tuning stages. During pretraining, we leverage the trained generator to directly create \text{M}odification \text{T}ext-oriented \text{S}ynthetic \text{T}riplets (MTST) conditioned on pairs of images. 
For fine-tuning, we first synthesize reverse modification text to connect the target image back to the reference image. Subsequently, we devise a two-hop alignment strategy to incrementally close the semantic gap between the multimodal pair and the target image. We initially learn an implicit prototype utilizing both the original triplet and its reversed version in a cycle manner, followed by combining the implicit prototype feature with the modification text to facilitate accurate alignment with the target image.  
Extensive experiments validate the efficacy of the generated triplets and confirm that our proposed methodology attains competitive recall on both the CIRR and FashionIQ benchmarks. 
\end{abstract}


\begin{figure}[!t]
  \centering
    \includegraphics[width=0.9\linewidth]{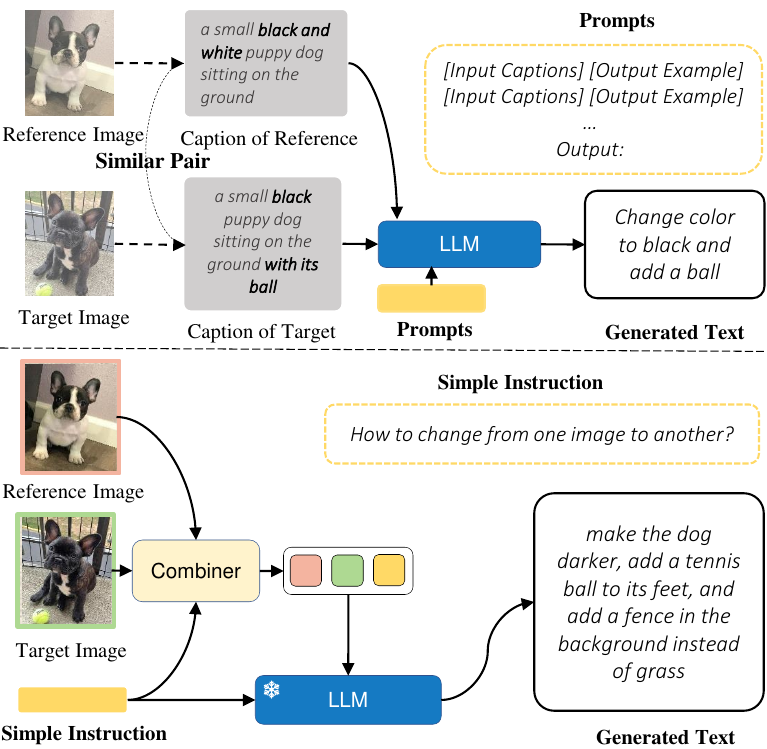}
  \caption{\textbf{Comparison of existing automatic modifier generation method and ours.} Above: Existing methods utilize indirect information, such as labels and captions of images to generate modifiers, which often results in poorer notation quality, limited text length, and a lack of diversity in textual form. Below: Our proposed method combines the image pair and instruction, mapping them into a frozen LLM for text generation. This allows for a more flexible description of the details between images and generates high-quality modification text with controllable length.}
  \label{fig:simple}
\vspace{-0.5cm}
\end{figure}

\section{Introduction}
\IEEEPARstart{I}{n} the context of Composed Image Retrieval (CIR), a given reference image and the modification text (also known as modifier) are utilized to amalgamate information across both visual and textual modalities, to pinpoint the most congruous target image within an image gallery. In contrast to the conventional image retrieval tasks reliant solely on textual information\cite{blip2,karthikeyan2014survey} or tag information \cite{tag}, the CIR model demands superior feature extraction, fusion, and inference capabilities. 
Images encapsulate rich visual information and intuitive perceptions, while text provides precise descriptions and semantic understanding of image content. The fusion of these two modalities for retrieval purposes can support the identification of the target image more accurately. As a result, CIR is regarded as a meaningful and promising research area and extensive efforts have been dedicated to this task \cite{cirr,tirg, joint, Comprehensive, fashion200k,mar,mcl,mmt,nsfse}. Common practices typically utilize the well-annotated triplets to train the models. However, the annotation of triplets requires remarkable human efforts and triplets cannot be collected from social web-like image-text pairs.

Motivated by the achievements of AIGC \cite{stablediffusion,dalle2,gpt4,instuctp2p,llama,baichuan}, a straightforward solution for this issue is the synthesis of triplets. Predominantly, the procedure can be bifurcated into two methodologies: \textbf{(1)} One might resort to leveraging image editing models \cite{instuctp2p} to manipulate the reference images based on the provided modification text. However, there is a domain gap between the real image and the synthesized one. \textbf{(2)} Some alternative approaches \cite{compodiff,dataroaming,covr} utilize labels or captions of two related images and combine that information with some delicate prompts or templates as input to large language models to generate modification texts indirectly (\emph{see Figure \ref{fig:simple}}). This approach often results in poorer notation quality, limited text length, and a lack of diversity in textual form. 

With the above considerations,  we resort to generating the modification text using the reference and target image as input. This solution, on the one hand, produces the textual modality, whose gap is smaller than the visual modality. On the other hand, sufficient triplets can be generated by simply feeding two images, which is similar to the manual labeling process. To generate reliable modification text, we leverage the annotated triplets as the training source and train a modification text generator by tuning the multimodal large-scale models. 
Subsequently, we mine a substantial number of relevant image pairs using similar labels or image sets.
With the trained generator and image pairs, we can freely generate the \textbf{M}odification \textbf{T}ext-oriented \textbf{S}ynthetic \textbf{T}riplets (MTST) to augment the original benchmarks, supporting a large scale pretraining.
\begin{figure}[t]
    \centering
    \includegraphics[width=0.9\linewidth]{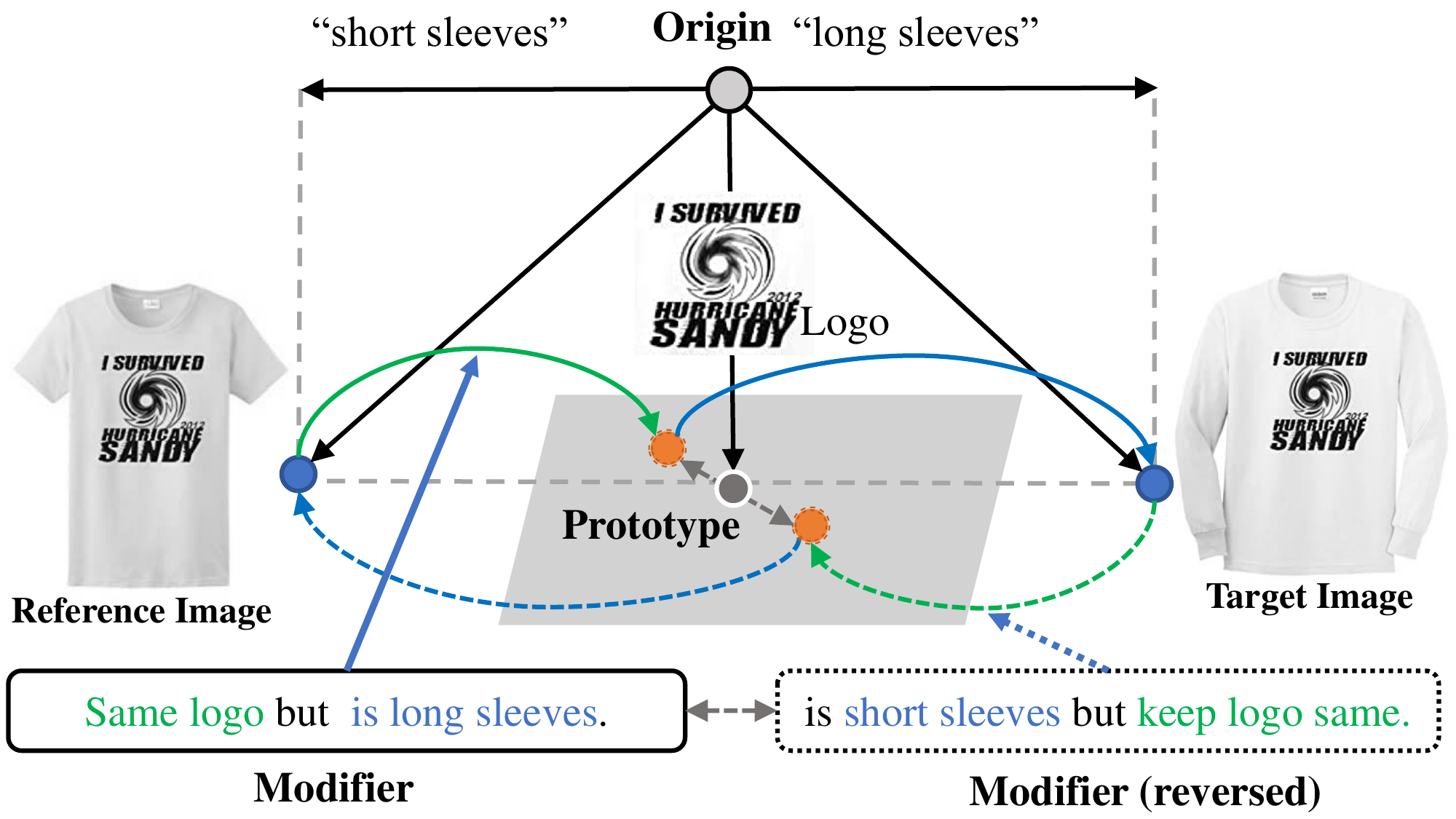}
    \caption{\textbf{An intuitive prototype example} of the reference image's prototype reached by the modifier and the target image's prototype reached by the reversed modifier. The \textcolor[HTML]{00b050}{green} dashed line represents the content preserved in the image based on the corresponding modification text. The \textcolor[HTML]{0070c0}{blue} dashed line represents the modifications and additions made to the prototype based on the corresponding modified text.}
    \label{fig:simpleexample}
    \vspace{-0.5cm}
  \end{figure}
The final model can be harvested by following the standard CIR learning protocol for fine-tuning, which merges query pairs and aligns the queries with target images. 

Nonetheless, a key distinction lies in the nature of the modification text in CIR, which acts less as a content descriptor and more akin to a two-part instructional guide. The information in the modification text for one reference image can be divided into two parts: one part pertains to the elements or features of the reference image that need to be preserved, which are usually expressed implicitly, and the other part includes new additions or changes.
Figure \ref{fig:simpleexample} shows one example to illustrate this characteristic of the modifier. If we start from the origin and add two orthogonal pieces of information, [logo] and ``short sleeves'' result in the reference image, while [logo] and ``long sleeves'' point to the target image. Therefore, the modifier, ``Same logo but is long sleeves'', which transitions from the reference to the target image, can be decoupled into two steps.
The first step is to retain identical or similar information from the reference image, corresponding to ``same logo'', along with other implicitly unmentioned details such as the collar style and color. 
We denote these preserved traits as the implicit prototype, generated based on the reference image and the corresponding modification text. The second step combines the implicit prototype with new or altered content from the modifier, such as ``longer sleeves'', to form a composite query to retrieve the most suitable target image. 

Motivated by this observation, we design a Prototypical Two-Hop Alignment (PTHA) strategy to progressively bridge the semantic gap between the multi-modal query and the target image. In particular, PTHA decouples the alignment into two steps: the first step generates the reverse modifier using the MTST generator and learns the implicit prototype preserving the sharing clues, while the second step combines the implicit prototype with the modifier to align with the target image.
In a nutshell, the contribution can be summarized as follows:
\begin{itemize}
  \item[$\bullet$] 
  Considering the triplet scarcity in composed image retrieval (CIR), we contribute a generation framework of Modification Text-oriented Synthetic Triplets (MTST) to augment the existing benchmarks with high-quality synthetic triplets, supporting effective pre-training for CIR.
  \item[$\bullet$] 

  We build two large-scale pretraining datasets for nature and fashion domains with our trained modification text generator, containing 800K and 580K triplets with expressive modification texts. 
  \item[$\bullet$] 
  A Prototypical Two-Hop Alignment (PTHA) strategy is proposed, which decouples the CIR problem as a two-step alignment paradigm to gradually bridge the gap between the multimodal query and the target image.
   \item[$\bullet$] Benefiting from the generated high-quality triplets and our devised PTHA network, we achieve comparable results, with improvements of {+2.39} in Avg. on the CIRR benchmark from nature and +1.57 in Avg. on the FashionIQ benchmark.

\end{itemize}

\section{Related Work}

\subsection{Vision Language Models}
\noindent In recent years, significant progress has been made in Vision Language Models(VLMs)\cite{clip,align,albef,blip,lin2024mope,lin2024duquant,doge,MM23,TMM23}. Typical models like CLIP~\cite{clip} and ALIGN \cite{align} achieve cross-modal understanding by leveraging contrastive learning on large-scale image and text pairs. Li \etal~\cite{albef} introduce image-text matching and masked language modeling (MLM) tasks during training to enhance fine-grained matching. BLIP~\cite{blip}  equip the pre-trained models with text generation capabilities by language modeling (LM). 
With a similar spirit, some recent works further fine-tune the cross-modality model for different downstream tasks, such as text-based person retrieval~\cite{yang2023towards} and drone localization~\cite{chu2025towards}.
The emergence of various Large Language Models (LLMs)~\cite{gpt4,llama,chain,vicuna} has also influenced the development of visual language models, as they possess vast knowledge and powerful text generation capabilities. 
LLaVa~\cite{llava} directly maps visual features to LLMs and aligns spaces through fintuning. 
BLIP2~\cite{blip2} establishes a bridge between vision language base models and various open-source LLMs by deploying a Q-Former on filtered data. 
InstructBLIP~\cite{instructblip} further improves performance using instruction tuning and exhibits enhanced text generation capabilities while reducing training costs through LLM freezing. 
We deploy instruction tuning to composed image retrieval and make it possible to generate modifiers using related images.

\subsection{Composed Image Retrieval}

\noindent Image retrieval is an important research task in multi-modal field. It aims to retrieve target images from a gallery based on a given query. This can be done by solely using text descriptions or by using images~\cite{deepfashion,sketch} to retrieve similar or related images. However, single-modal tasks such as text-based image retrieval or image-based image retrieval cannot accurately and conveniently meet the specific retrieval needs of certain scenarios. To address this issue, the Composed Image Retrieval task has been proposed\cite{cirr,tirg,Comprehensive,fashioniq}, which involves integrating the reference image feature and supplementing or modifying the textual feature to retrieve the target image.
There have been efforts in training lightweight connection layers to obtain fused features from image and text representations. ARTEMIS~\cite{atermis} combines triplets through explicit matching and implicit similarity, and Baldrati \etal~\cite{clip4cir2} proposes a combiner to leverage CLIP visual and textual representations. Liu \etal~\cite{rerank} proposes a re-ranking method after the first selection. In very recent works, {many existing works\cite{circo,gu2024lincir,pic2word,vaze2023gen,magiclens} leverage large amounts of external data to achieve zero-shot CIR capabilities. MagicLens\cite{magiclens} achieves strong performance in zero-shot CIR while also making progress in richer relations beyond image similarity.} SPRC~\cite{sprc} utilizes Q-Former to extract sentence-level prompts and guides sentence-level prompt generation aligned with an auxiliary text prompt. In addition, there have been works that enhance task performance by introducing additional datasets for pre-training~\cite{covr,dataroaming,compodiff}. These datasets provide extra training examples and diverse data distributions, allowing the models to learn more comprehensive and robust representations. 
In our work, we design a prototypical two-hop alignment network to decompose CIR into an implicit prototype learning module and fusion module. In the implicit prototype learning module, we utilize generated reversed modifiers to benefit implicit prototype learning.
\subsection{Composed Image Retrieval Triplet Generation}
\noindent
In previous works, CIR triplet generation has been primarily achieved through manual and automatic methods: 
\textbf{Manual Annotated.}
The CIRR dataset~\cite{cirr} consists of manually annotated textual triplets, which are derived from a subset of images from the NLVR2 dataset~\cite{nlvr2}, representing real-world scenarios. The FashionIQ dataset~\cite{fashioniq} comprises manually selected pairs of similar fashion images, along with human-annotated textual triplets, specifically curated for the fashion domain. 
\textbf{Automatic Annotated.}
Han \etal~\cite{fashion200k} employed the differences in manually annotated attribute labels of fashion200k dataset images to generate modified texts using a triplet template. In recent years, there have been proposed methods that leverage automatic generation techniques from other tasks and models. LaSCo~\cite{covr} utilized VQA 2.0~\cite{vqa2} to construct triplets by using different answers for similar images and the same question, employing GPT 3.0~\cite{gpt3}, followed by manual quality control. CompoDiff~\cite{compodiff} built triplets based on {InstructPix2Pix}\cite{instuctp2p}, using text descriptions collected from both human annotators and large-scale model generation and generating images using Stable Diffusion\cite{stablediffusion}. CoVR~\cite{covr} employed similar video captions to filter similar image pairs and trained an MTG-LLM to generate a modifier using two similar captions, forming triplets.
Compared to existing methods, we incorporate images into the training of modifier generation and map visual features into the space of a large model. We propose a lightweight text generation method that is more flexible, diverse, and controllable in length while maintaining low training costs.

\section{\text{M}odification \text{T}ext-oriented \text{S}ynthetic \text{T}riplets (MTST) Generation}
\label{sec:MTSTGeneration}
\begin{figure*}[!t]
  \centering
  \includegraphics[width=0.9\linewidth]{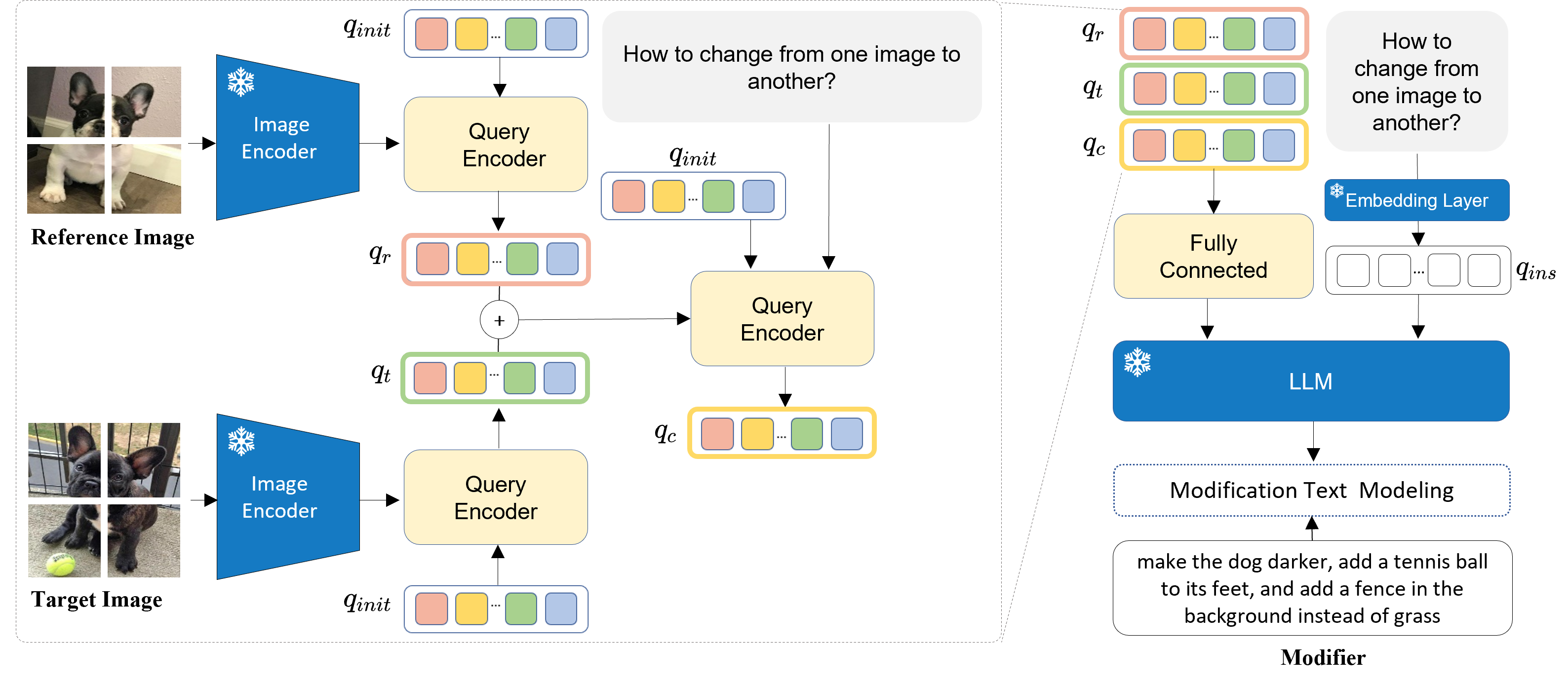}
    \vspace{-0.4cm}
  \caption{\textbf{An overview of MTST Generator architecture.} The paired reference images and target images are fed into an Image Encoder and a trainable Query Encoder to get their respective features ${q_r,q_t}$. These representations are then concatenated and fused with the instruction using the same Query Encoder to obtain a fusion representation ${q_c}$. Representations ${q_r,q_t,q_c}$ along with the instruction embedding are concatenated and fed into a frozen LLM to generate modification text.
  }
  \label{fig:arch1}
\vspace{-0.5cm}
\end{figure*}

\subsection{MTST Generator}
\label{sec:mtst}
\subsubsection{Architecture}

\noindent \text{M}odification \text{T}ext-oriented \text{S}ynthetic \text{T}riplets (MTST) generator takes the reference and target images as input and outputs the modification text. To produce high-quality text, we follow the multimodal large model~\cite{blip2,instructblip} to design our MTST generator, \emph{i.e.,} image encoder extracts the image features and a Large Language Model (LLM) is followed to give reliable text output. Figure \ref{fig:arch1} depicts the overall architecture, consisting of an Image Encoder and a Query Encoder for vision feature extraction, a Fully Connected Layer bridging the vision feature to the LLM, and a LLM as the text prediction module. We keep the Image Encoder and LLM frozen and only train the Query Encoder and Fully Connected Layer to map the learned prompt feature to LLM space.

\subsubsection{Generator Training} 
\noindent The manually annotated (training) triplets in the standard benchmark like CIRR~\cite{cirr} are taken as the training samples. To guide the network learning, we also provide an instruction, ``\texttt{How to change from one image to another?}'', which is incorporated into the Query Encoder to clarify the purpose.   
In specific, the reference image and target image are separately input to the Image Encoder and Query Encoder to get their respective query features ${q_r,q_t}$ from initial query tokens ${q_{init}}$.

To acquire the task-oriented features, we next feed the instruction and vision features ${q_r,q_t}$ into Query Encoder to obtain a composed representation ${q_c}$. 
Subsequently, we pass ${q_r}$, ${q_t}$, and ${q_c}$ through the projection layer ${FC_{llm}}$ to produce a comprehensive vision feature, which is then concatenated with the instruction embedding ${q_{ins}}$ and fed into LLM for text generation:
\begin{equation}
\text{input}_{llm}=FC_{llm}(q_r\oplus q_t \oplus q_c)\oplus q_{ins},
\label{eq:3} 
\end{equation}

For modification text modeling, we deploy a language generative task auto-regressively to predict the next token of the modification text by maximizing the conditional likelihood:
\begin{equation}
\begin{split}
\mathcal{L}_{gen} = -\mathbb{E}_{(I_r,I_t,t)\sim\mathcal{D}}[\sum_{m=1}^{M} logP(t_m|\text{input}_{llm},t_0,\cdots,t_{m-1})],
\label{eq:4} 
\end{split}
\end{equation}
where $M$ denotes the length of modification text $t$ and $t_m$ denotes the $m^{th}$ token of t.  $\mathcal{D}$ is the distribution of triplets. ${I_r}$ and ${I_t}$ denote the reference and the target image, respectively.

\subsection{Grounded MTST Generation}
\label{sec:Groundedmtst}
\noindent In this paper, we focus on two common domains, \emph{i.e.,} the nature and the fashion, for MTST generation. We take the popular CIRR and FahsionIQ benchmarks as the source data.
To generate triplets resembling the real world, we carefully design the image sampling strategies for the triplet generation.

 \textbf{${\text{CIRR}_\text{MTST}}$.} We employ two strategies for selecting image pairs from the CIRR dataset: (1) The CIRR training dataset comprises 3,345 image sets, each featuring six analogous images, from which the training triplets are sampled. Pairwise combinations of images within these clusters yield a total of 100,350 unique pairs. (2) We create new image sets by combining images with the same category from the NLVR2 dataset\cite{nlvr2}, which is the source dataset for CIRR images. We then pair these newly created image sets, resulting in 707,745 image pairs.
Therefore, by extending MTST on these image pairs, we generate a total of 808,095 triplets on the CIRR dataset.

\textbf{${\text{FashionIQ}_\text{MTST}}$} For the FashionIQ dataset, each image has multiple labels that describe its style, such as `short sleeves', `v-neck', `hoodie'. We classify the images based on their labels, and images with the same label, share certain characteristics and similarities. From these images, we can select the image pairs we need.
However, some labels usually correspond to a large number of images. For example, the label ``long sleeve'' corresponds to 1,006 images. If we pair the images with same label all together, we would end up with 1,011,030 image pairs for only one category. This could result in weak image correlations and an imbalanced dataset. To address this, we impose a limit to the number of image pairs, ensuring it does not exceed three times the number of images in its respective category. As a result, we generated a total of 579,114 triplets from the dress, shirt, and toptee categories in FashionIQ. Table ~\ref{tab:dataset} presents the statistics of our final datasets and the comparison with the existing CIR datasets. Our proposed MTST exhibits several salient advantages: 

\begin{itemize}

\item[$\bullet$] \textbf{Narrower domain gap to the annotated triplets.} By resorting to text generation, we effectively mitigate domain discrepancies compared to image generation methods. Furthermore, benefiting from the paradigm of directly utilizing real images as input for text generation, we bypass visual domain gaps inherent in other approaches. 

\item[$\bullet$] \textbf{Expressive modification text}. Leveraging the capabilities of Large Language Models (LLMs), MTST yields modification texts that are more verbose than those found in existing benchmarks. This characteristic allows for richer and more expressive content representation.    

\item[$\bullet$] \textbf{Greater flexibility in triplet generation.}  The MTST framework necessitates only two images for generating corresponding text, thus demonstrating a high degree of flexibility. This enables large-scale triplet generation without compromising efficiency or diversity.  

\end{itemize}
\begin{table*}[t]
    \caption{
    \textbf{Statistics of existing CIR datasets and our generated dataset:}  We expand the triplets of CIRR and FashionIQ datasets using our MTST generator. The table compares the number of triplets, unique images, unique words, and the average length of modification text. 
          }
    \centering
    \vspace{-0.2cm}
    \resizebox{0.9\linewidth}{!}{%
    \centering
    \begin{tabular}{l|cc|c c c c }
        \toprule
        \textbf{Name}  &\textbf{Domain} 
 &\textbf{Image Source} 
& \textbf{\#Triplets} & \textbf{\#Unique images} & \textbf{\#Average length} & \textbf{\#Unique Words} \\
        \midrule
        FashionIQ (train)\cite{fashioniq}  &Fashion 
 &FashionIQ (train) 
& 16,914 & 23,813 & 54.90 & 4,253 \\
        CIRR (train)\cite{cirr}  &Nature 
 &CIRR (train) 
& 36,761 & 21,185 & 59.51 & 7,129 \\ \midrule
        SynthTriplets 18M\cite{compodiff} &Nature
 &Synthetic
&18,000,000&-&-&-\\
        LaSCo\cite{dataroaming} &Nature 
 &VQA2.0\cite{vqa2} 
& 389,305 & 121,479 & 30.7 & 13,488 \\ 
        WebVid-CoVR\cite{covr}  &Nature 
 &WebVid2M,WebVid10M\cite{webvids} 
& 1,648,789 & 130,775 & 23.36 & 19,163 \\ \midrule
\rowcolor{orange!10}
        \textbf{${\text{FashionIQ}_\text{MTST}}$}&Fashion 
 &FashionIQ\cite{fashioniq} (train) 
& 579,114 &26,048 & \textbf{61.66} & \textbf{5,212} \\
\rowcolor{orange!10}
        \textbf{${\text{CIRR}_\text{MTST}}$}&Nature  &NLVR2\cite{nlvr2} (train) & 808,096 & 103,170 & \textbf{113.03} & \textbf{19,681} \\ 
        \bottomrule
    \end{tabular}}
    \label{tab:dataset}
    \vspace{-0.5cm}
\end{table*}

\section{Prototypical Two-Hop Alignment Network}
\label{sec:Prototypical}
\begin{figure*}[!t]
  \centering
  \includegraphics[width=0.9\linewidth]{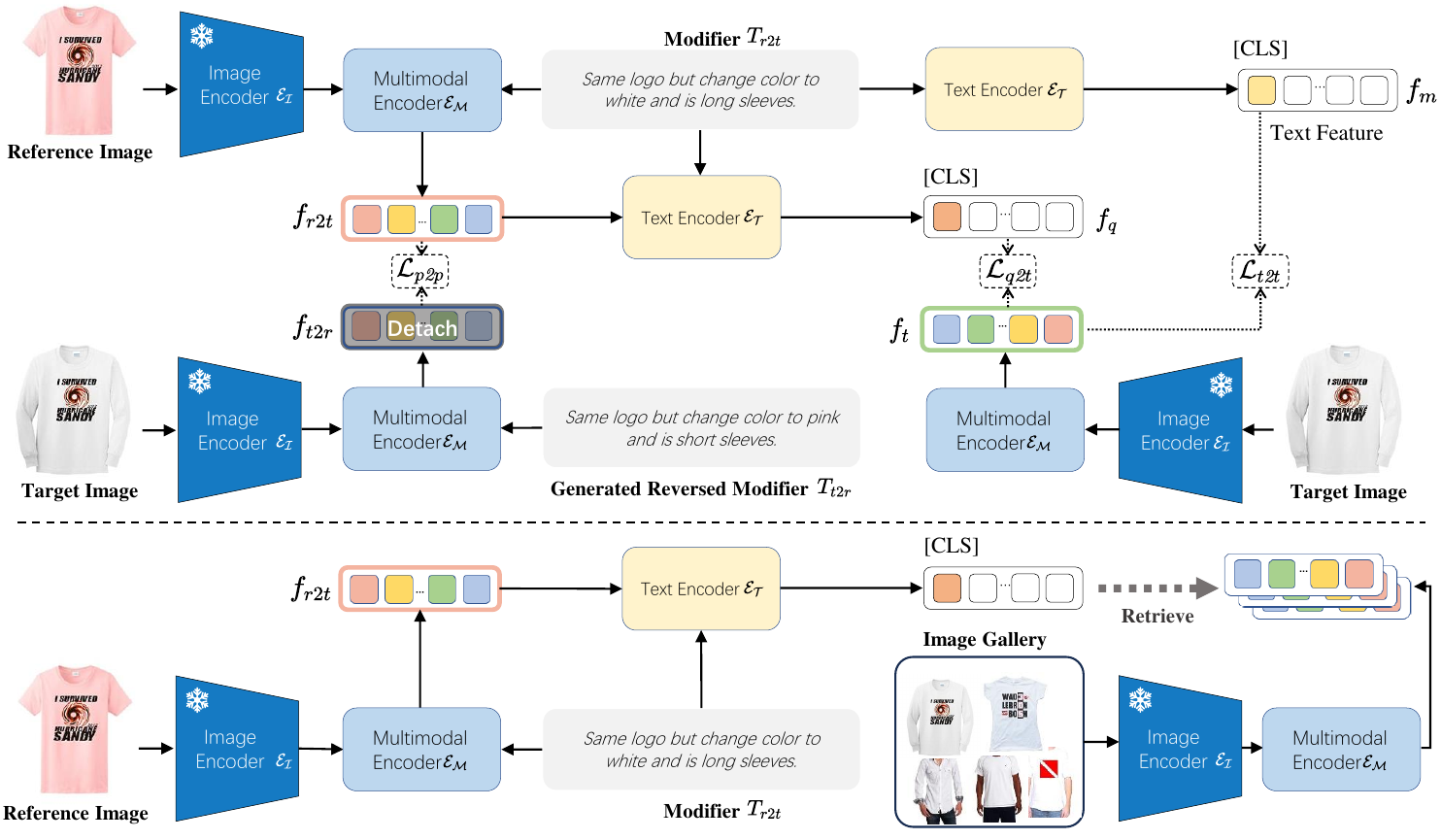}
    \vspace{-0.2cm}
  \caption{\textbf{Top:} Overview of the training phase. We first employ implicit prototype learning between ${f_{r2t}}$ and detached ${f_{t2r}}$. ${f_{t2r}}$ is extracted from the target image and generated reversed text using the MTST generator. In the second step, we utilize contrastive loss between fusion feature $f_q$ and target image  ${f_t}$, text-only feature ${f_m}$, and target image feature ${f_t}$. \textbf{Bottom:} Overview of the inference phase. We leverage the fusion feature $f_q$ to compute similarity with the features extracted from the image gallery to perform retrieval.}
  \label{fig:method}
  \vspace{-0.5cm}
\end{figure*}

\noindent As illustrated in Figure \ref{fig:method}, our prototypical two-hop alignment (PTHA) network comprises an image encoder and a text encoder for image and text feature extraction respectively, and a multimodal encoder to combine the multimodal query pair. During training, we first generate the reversed modifier using MTST and then apply the proposed PTHA to learn the network. When inference, we utilize the fusion feature to compute cosine similarity with the image feature extracted from candidates in the image gallery to perform retrieval.

\subsection{Pre-training with MTST}
\label{pretrain}
\noindent Before optimizing our PTHA network, we first adopt MTST to perform pre-training, pursuing a better initialization for the subsequent learning.  
Formally, we first encode the reference image $I_r$ adpoting a frozen image encoder $\mathcal{E_I}$, and fuse the resulted representation with modification text ${T_{r2t}}$ with the multi-modal encoder${\mathcal{E_M}}$:
\begin{equation}
f_{r2t} = \mathcal{E_M} (\mathcal{E_I}(I_r),T_{r2t}).
\label{eq:feat}
\end{equation}

Following SPRC~\cite{sprc}, we take the multimodal feature $f_{r2t}$ as the textual prompt, which is then fed into the text encoder $\mathcal{E_T}$ with modification text to produce the fusion feature of the reference image and modification text:
\begin{equation}
f_q = \mathcal{E_T}(f_{r2t},T_{r2t}).
\end{equation}

We utilize the same Image Encoder $\mathcal{E_I}$ to encode target image $I_t$ and the same multi-modal encoder ${\mathcal{E_M}}$ to produce the target feature of query pair:
\begin{equation}
f_t = \mathcal{E_M}(\mathcal{E_I}(I_t)).
\end{equation}

Subsequently, we deploy contrastive learning loss query feature $f_q$ and target image feature $f_t$ to train the network:
\begin{equation}
\label{eq:p2t} 
\mathcal{L}_{q2t} = -\frac{1}{\mathcal{|B|}} \sum_{i=1}^{ \mathcal{|B|}  } \log \frac{\exp(\text{sim}(f_{q_i} , f_{t_i})   / \tau)}{\sum_{j=1}^{\mathcal{|B|}} \exp(\text{sim}(f_{q_i} , f_{t_j}) / \tau)},
\end{equation}
where ${\mathcal{B}}$ denotes the input batch, ${f_{q_i}}$ and ${f_{t_i}}$ denotes the $i$-th fusion feature and target feature in ${\mathcal{B}}$ respectively, and ${\tau}$ is a learnable temperature parameter. 
For similarity calculation, we adopt the \texttt{[CLS]} token in $f_q$, \ie ${f_{q_{cls}}}$, to query the target image embeddings and take the max pool as a similarity estimation:
\begin{equation}
\label{eq:p2t2} 
    \text{sim}(f_q, f_t) = \max_{k\in[1,N]} \frac{f_{q_{cls}}\cdot f_{t}[k]}{||f_{q_{cls}}||\cdot||f_{t}[k]||},
\end{equation}
where $f_{t}[k]$ means the $k$-th token embedding in $f_t$.
\subsection{Prototype-bridged Two-Hop Fine-Tuning}

\noindent Motivated by the intuition in Figure \ref{fig:simpleexample}, we perform a two-hop alignment stategy during the fine-tuning phase.
\textbf{Implicit Prototype Learning via Reverse Text.} In the first step, we target to learn the implicit prototype preserving the shared information between the images. Particularly, we force the two images to approach each other via the respective text guidance, to reach a feature (implicit prototype) containing shared information in the two images. To support this, we first synthesize the modifiers from the target image to the reference image adopting the trained MTST generator. 

In specific, let ${T_{t2r}}$ be the reverse modification text, then similar to Eq. \ref{eq:feat}, the representation ${f_{t2r}}$ of the reversed pair can be obtained from the target image ${I_t}$ and the generated reverse modification text ${T_{t2r}}$.

Subsequently, we learn the implicit prototype by making the two multi-modal features step towards each other:
\begin{equation}
\begin{split}
\mathcal{L}_{p2p}= \frac{1}{\mathcal{|B|}}\sum_{i=1}^{\mathcal{|B|}} (\overline{f}_{r2t_i} - \overline{f}_{t2r_i} )^2,
\label{eq:6} 
\end{split}
\end{equation}

where $\overline{f}$ denotes the mean of $N$ tokens of feature $f$. During training, we stop the gradient propagation of ${f_{t2r}}$. The gradient update is halted for the reversed multimodal pair due to two considerations: (1) The reversed text is only generated during training for $L_{p2p}$ calculation, consequently, we maintain the integrity of implicit prototype details in the real query pair's features to align across both training and inference stages; (2) Mitigating the likelihood of model degeneration or collapse.

\noindent\textbf{Implicit Prototype-bridge Alignment.} In the second step, we further fuse the learned implicit prototype ${f_{r2t}}$ with the modifier ${T}$ to combine the necessary modifications or additions, yielding the composite feature $f_q$ for retrieval, which is then aligned with the target image with a contrastive procedure similar to \cref{eq:p2t}. 
Besides, following SPRC\cite{sprc}, we also align the modification text and the target image as an auxiliary constraint $\mathcal{L}_{t2t}$, which matches the content in the modification text feature $f_m$ to the target image feature $f_t$ to aid the learning of dominated constraints, on the other hand, narrows the semantic gap between the modification text and the target image, such that ease the alignment of query pair and target image in feature space. The constraint is also specified as a contrastive procedure similar to \cref{eq:p2t} with the features of the target image and the modification text.
Our final objective is formulated as:
\begin{equation}
\label{finaloss}
\mathcal{L} = \mathcal{L}_{q2t} +\mathcal{L}_{t2t}+\alpha\mathcal{L}_{p2p},
\end{equation}
where $\alpha$ is a non-negative trade-off hyper-parameter.

\section{Experiment}
\subsection{Experimental Setup}
\subsubsection{Datasets and Evaluation Metrics}
Following previous work\cite{clip4cir2,blip4cir,sprc,spn}, the real-world dataset CIRR and the fashion domain dataset FashionIQ are considered. Both are manually annotated CIR datasets based on real images. CIRR is created from 21k real-world images from NLVR2\cite{nlvr2}, producing 36k triples.
FashionIQ primarily contains images from three categories in the fashion domain: dress, shirt, and toptee. It comprises a total of 77k fashion product images and 30k triplets, each pair of images contains two different relative captions annotated by two individuals.
Following the metric settings of the standard evaluation experiment, we report the average recall at rank K (R@K).
For FashionIQ, we report the R@10 and R@50 on the val set in three categories. For CIRR, we disclose the R@1, 5, 10, and 50 as well as $\text{Recall}_\text{subset}\text{@}$1, 2, 3 on test set. $\text{Recall}_\text{subset}\text{@K}$ serves as a benchmark for fine-grained matching, considering each image set composed of 6 similar images as the search space.

\subsubsection{Implementation Details}
{
{\bf MTST Generator.} The MTST Generator is fine-tuned on the CIR task based on the InstructBlip-Vicuna-7b\cite{instructblip} base model, with only the Q-Former fine-tuned. The number of query tokens in the Q-Former \cite{blip2} is set to 32. The frozen vicuna checkpoint is specified as vicuna-7b-v1.1 \cite{vicuna}. The model weights are loaded from a pre-trained model available from InstructBLIP \cite{instructblip}.
}

{
{\bf PTHA.} The model architecture in PTHA is the same as SPRC~\cite{sprc}. The multimodal encoder is the query encoder in BLIP-2\cite{blip2}, the text encoder is the text encoder in BLIP-2.  The vision encoder is the frozen ViT-g/14 from EVA-CLIP\cite{eva}. We employ two-stage strategy. The batch size is set to 128 and the number N of query tokens is set to 32 for both stages. The initial model weights of pretraining is from BLIP2-pretrained~\cite{blip2}. In the fine-tuning stage, we follow all the settings of SPRC~\cite{sprc}. During our finetuning stage on CIRR after pretraining on ${\text{CIRR}_\text{MTST}}$, we replace the \texttt{[CLS]} token ${f_{q_{cls}}}$ with the average embedding of N query tokens and the \texttt{[CLS]} token in $f_q$, \ie ${f_{q_{avg}}}$ in \cref{eq:p2t2}.}

\vspace{-0.4cm}
\subsection{Quantitative Results}
\begin{table*}[t]
\caption{\textbf{Comparison on CIRR test set.} "Pretraning data" is the dataset for model pretraining. "Avg." means (Recall@5 + $\text{Recall}_\text{subset}\text{@1}$)/2. The best result is indicated in \textbf{bold}, while the second best is \underline{underlined}. Our proposed PTHA with pretraining on ${\text{CIRR}_\text{MTST}}$ outperforms the previous method in all metrics. * indicates that the method deploys an extra re-ranking strategy.{† indicates the utilizing of extra training method SPN4CIR\cite{spn}.}}
\label{tab:cirrresult}
\centering
\vspace{-0.2cm}

\resizebox{1\linewidth}{!}{
\begin{tabular}{lll|cccc|ccc|c}
\toprule
 && &  \multicolumn{4}{c|}{\textbf{$\text{Recall@K}$}} &
  \multicolumn{3}{c|}{\textbf{$\text{Recall}_\text{subset}\text{@K}$}} &{\textbf{Avg.}}
 \\\cmidrule(lr){4-7}\cmidrule(lr){8-10}

          {\textbf{Method}} &backbone &\textbf{Pretraining Data}&\textbf{K=1}&\textbf{K=5}&\textbf{K=10}&\textbf{K=50}&\textbf{K=1}&\textbf{K=2}&\textbf{K=3} & \\ 
\midrule
MAAF~\cite{maaf}  &w/o VLP   & - &    10.31  &   33.03   &  48.30    & 80.06     & 21.05     &  41.81    & 61.60     & 27.04       \\
TIRG~\cite{tirg}   &w/o VLP   & - &     14.61 & 48.37     & 64.08     &   90.03   &  22.67    &  44.97    & 65.14     & 35.52       \\

ARTEMIS~\cite{atermis} &w/o VLP  & - & 16.96     &  46.10    &  61.31    &  87.73    & 39.99     & 62.20     & 75.67     & 43.05      \\
CIRPLANT w/OSCAR~\cite{cirr} &w/o VLP & - &   19.55   &   52.55   &  68.39    &    92.38  & 39.20     & 63.03     & 79.49     & 45.88       \\
ComqueryFormer\cite{mmt} &CLIP& - &   25.76   &   61.76   &  75.90   &    95.13  & 51.86    & 76.26    &89.25   & 56.81  \\
NSFSE~\cite{nsfse} &CLIP& - &   20.70   &   52.50   &  67.96    &    90.74  & 44.20     & 65.53     & 78.50     & 48.35       \\

Compdiff~\cite{compodiff}& CLIP  & SynthTriplets &    22.35  &   54.36   & 73.41     & 91.77     &  35.84    &  56.11    &  76.60    & 29.10       \\
CLIP4CIR~\cite{clip4cir2}  &CLIP & - &  38.53    &  69.98    &   81.86   &  95.93    &  68.19    & 85.64     & 94.17     &     69.09   \\
BLIP4CIR+Bi~\cite{blip4cir} &BLIP & - &  40.15    &  73.08    &  83.88    &   96.27   & 72.10     & 88.27     & 95.93     & 72.59       \\
CASE~\cite{dataroaming} &BLIP       & LaSCo &  49.35    &    80.02  &   88.75   &   97.47   &   76.48   &  90.37    &  95.71    &    78.25    \\
CoVR-BLIP~\cite{covr}&BLIP  & {WebVid-CoVR} & 49.69&   78.60   &    86.77  &   94.31   &    75.01  &   88.12   &    93.16  &   80.81     \\
Reranking*~\cite{rerank} &BLIP & - &    50.55  &   81.75   &    89.78  &   97.18   &   80.04   &   91.90   &    96.58  &   80.90     \\

SPRC ~\cite{sprc}  &BLIP-2   & - &   51.96   &   82.12   &   89.74   &   97.69   &  80.65   &    92.31  &  96.60    &    81.39   \\
SPRC${^2}$*\cite{sprc}  &BLIP-2     & - &  54.15 & {83.01} & {90.39} & {98.17} & \textbf{82.31} & {92.68} & {96.87} & {82.66} \\
SPRC†\cite{sprc}\cite{spn}&BLIP-2     & - &  \underline{55.06} & {83.83} & {90.87} & \underline{98.29} & {81.54} & {92.65} & {97.04} & {82.69}
 \\

\midrule
 Baseline  &BLIP-2    & - &    51.39  &   81.95   &   89.92   &   97.90   &   78.98   &   91.78  &   96.36  &   80.46    \\
 
PTHA (Ours) &BLIP-2     & - &    51.85  &   82.1   &   89.93   &   97.98   &   80.32   &   92.36  &   96.70  &   81.21    \\
PTHA (Ours)    &BLIP-2     & {${\text{CIRR}_\text{MTST}}$}&  {54.70}   &     \underline{84.05}&\underline{90.89}&   {98.26}   &  {81.64}    &   \underline{93.30}   &  \underline{97.30}   &   \underline{82.85}\\
\rowcolor{orange!10}
PTHA (Ours)†~\cite{spn}   &BLIP-2     & \textbf{${\text{CIRR}_\text{MTST}}$}&  \textbf{56.43}   &     \textbf{84.92}&\textbf{91.74}&   \textbf{98.43}   &  \underline{82.12}    &   \textbf{93.35}   &  \textbf{97.42}   &   \textbf{83.52}\\
\bottomrule
\end{tabular}
}
\vspace{-0.3cm}
\end{table*}

\subsubsection{CIRR.} Table ~\ref{tab:cirrresult} shows the comparison result on CIRR. It is worth mentioning that the results of pretraining on our ${\text{CIRR}_\text{MTST}}$ and fine-tuning using our proposed PTHA framework outperforms all existing methods across all metrics. Compared to the SOTA model without an extra re-ranking strategy, \ie SPRC, we achieve improvements of {\textbf{+2.74}, \textbf{+1.93}, \textbf{+1.15}, and \textbf{+0.57}} in Recall@1, 5, 10, and 50, respectively, and obtain an overall average improvement of \textbf{+1.46}. 
{Additionally, we further fine-tuned our method using SPN4CIR\cite{spn}, achieving better performance. Compared to SPRC with SPN4CIR\cite{spn}, we achieve an improvement of +0.83 on average.}

\begin{table*}[t]
\centering
\caption{\textbf{Comparison on FashionIQ validation set.} ``Avg.'' means (Recall@10 + Recall@50)/2. * indicates that the method deploys extra re-ranking strategy.
}
\vspace{-0.2cm}
\label{tab:fiqresult}
\resizebox{1\linewidth}{!}{
\begin{tabular}{lll|cc|cc|cc|cc|c}
\toprule
 &
   &&
  \multicolumn{2}{c|}{\textbf{Dress}} &
  \multicolumn{2}{c|}{\textbf{Shirt}} &
  \multicolumn{2}{c|}{\textbf{Toptee}} &
  \multicolumn{2}{c|}{\textbf{Average}} &
 {\textbf{Avg.}} \\ \cmidrule(lr){4-5}\cmidrule(lr){6-7}\cmidrule(lr){8-9}\cmidrule(lr){10-11}
          {\textbf{Method}} &baseline& {\textbf{Pretraining Data}} & \textbf{R@10} & \textbf{R@50} & \textbf{R@10} & \textbf{R@50} & \textbf{R@10} & \textbf{R@50} & \textbf{R@10} & \textbf{R@50} &  \\ \midrule
    
TIRG\cite{tirg} & w/o VLP & - &     14.87 &  34.66    &   18.26   &  37.89    &  19.08    & 39.62     &   17.40   &  37.39    & 27.45 \\
CIRPLANT w/OSCAR~\cite{cirr}& w/o VLP  & - &     17.45 &   40.41   &    17.53  &   38.81   &    61.64  &    45.38  &   18.87   &   41.53   &  30.20\\
MAAF\cite{maaf}     & w/o VLP & - &      23.8&   48.6   &   21.3   &   44.2   &    27.9  &   53.6   &   24.3   &   48.8   & 36.6 \\
CurlingNet\cite{curling}& w/o VLP& - &    26.15  &  53.24    &   21.45   &  44.56    &  30.12    &  55.23    & 25.90     & 51.01     & 34.36 \\

CosMo\cite{cosmo}& w/o VLP   & - &     25.64 & 50.30     & 24.90     & 49.18     & 29.21     & 57.46     &  26.58    & 52.31     & 39.45 \\
ARTEMIS~\cite{atermis}& w/o VLP   & - &     25.68 &   51.25   &  28.59    &  55.06    &  21.57    & 44.13     & 25.25     &  50.08    & 37.67 \\

NSFSE~\cite{nsfse} & CLIP& - &      31.12  &  55.73   & 24.58     & 45.85     &  31.93    &  58.37    &  29.17    & 53.24     &  41.26\\
MUR\cite{chen2024composed} & CLIP & - &  32.61& 61.34 &33.23 &62.55 &41.40& 72.51 &35.75 &65.47&50.61\\
Css-Net\cite{css} & CLIP & - &  33.65& 63.16 &35.96 &61.96 &42.65& 70.70 &37.42 &65.27&51.35\\
CLIP4CIR\cite{clip4cir2}  & CLIP & - &      33.81& 59.40     & 39.99     & 60.45     &  41.41    &  65.37    &  38.82    & 61.74     &  50.03\\
ComqueryFormer\cite{mmt} & CLIP & - &  33.86& 61.08 &35.57 &62.19 &42.07& 69.30 &37.17 &64.19&50.68\\
CompoDiff~\cite{compodiff}   & CLIP    & SynthTriplets &   40.88   &   53.06   &   35.53   &   49.56   &  41.15    &  54.12    &   39.05   &  52.34    & 46.31 \\
FAME-VIL\cite{famevil}  & CLIP& -&     42.19 &   67.38   &   47.64   &   68.79   &   50.69   &   73.07   &   46.84   &   69.75   &  58.29\\
BLIP4CIR+Bi\cite{blip4cir}  & BLIP  & - &    42.09  &    67.33  &    41.76  &   64.28   &   46.61   &    70.32  &   43.49   &   67.31   & 55.40 \\

CoVR-BLIP\cite{covr}   & BLIP  & WebVid-CoVR & 44.55    & 69.03     & 48.43     & 67.42     & 52.60     & 74.31     & 48.53     & 70.25     &  59.39\\

CASE\cite{dataroaming} & BLIP     & LaSCo &  47.77    &   69.36   &   48.48   &   70.23   &   50.18   &  72.24    &   48.79   &   70.68   & 59.74 \\

Reranking*~\cite{rerank}& BLIP & - &     48.14 &     71.34 &   50.15   &  71.25    &    55.23  &   76.80   &    51.17  &   73.13   &  62.15\\
SPRC\cite{sprc}   & BLIP2    & - &  49.18    &  72.43    &   \underline{55.64}   &   73.89   &     \textbf{59.35}  &   78.58   &   \underline{54.92}   &   74.97   & \underline{64.85} \\

\midrule
 Baseline   & BLIP2  &-  &   48.04   & 72.65     & 53.54     & 73.91     &   57.37   &  78.85    & 52.98     &  75.13    & 64.07 \\

PTHA (Ours)    & BLIP2  &-  &   \underline{49.54}   & \underline{72.81}     & 55.50     & \underline{73.97}     &   57.96   &  \underline{78.96}    & 54.33     &  \underline{75.24}    & 64.79 \\
\rowcolor{orange!10}
PTHA (Ours)   & BLIP2     &\textbf{${\text{FashionIQ}_\text{MTST}}$}& \textbf{50.77}    &   \textbf{73.78}   &   \textbf{55.91}   &   \textbf{75.36}   &  \underline{58.28}   &   \textbf{79.70 }  &  \textbf{54.99}    &   \textbf{76.28}   &  \textbf{65.64}\\
\bottomrule
\end{tabular}
}
\vspace{-0.5cm}
\end{table*}

\subsubsection{FashinIQ.} We then evaluate our ${\text{FashionIQ}_\text{MTST}}$  and PTHA on FashionIQ. As shown in Table ~\ref{tab:fiqresult}, we observe a similar upward trend in performance improvement as it in CIRR, indicating consistent progress. Apart from the R@10 metric on Toptee, we surpass the second-best method, \ie SPRC in all other metrics, achieving an average improvement of \textbf{+0.79}. It indicates that our method and pretraining strategy are similarly applicable in the FashionIQ dataset.

\begin{table*}[t]
    \caption{\textbf{Ablation studies.} (a): Comparison of different loss combinations on Recall@5 and ${\text{Recall}_\text{subset}}$@1 metrics of CIRR validation set. ``\checkmark'' denotes the loss in the column is applied. We report the results on CIRR validation set. (b): Performance comparison of four methods w/ and w/o pre-training using ${\text{CIRR}_\text{MTST}}$700k. MTST pre-training brings clear performance improvements across all four methods. (c): Comparison of the results using original text and generated text for triplets. (d): Comparison of the results of PTHA and SPRC\cite{sprc} on CIRR val dataset. (e): We achieve admirable performance after first phase's pre-training on ${\text{CIRR}_\text{MTST}\\700K}$ by only utilizing a simple contrastive learning loss ${\mathcal{L}_{q2t} }$. {(f):Zero-shot CIR performance comparison on CIRCO [54] test set.}}
    \vspace{-0.2cm}

    \begin{minipage}[t]{1\linewidth}
        \subfloat[][{Ablation of Loss Functions and Pretraining}]{
        \resizebox{0.48\linewidth}{!}{
        \begin{tabular}{@{}l|ccc|cc|c@{}}
        \toprule
         &\multicolumn{3}{c|}{Losses}  & \multirow{2}*{$\text{Recall@5}$} & \multirow{2}*{$\text{Recall}_\text{subset}\text{@1}$} & \multirow{2}*{Avg. } 
         \\ 
         \cmidrule(lr){2-4}
            \multirow{6}*{\shortstack{w/o \\ pretrain}}   &$\mathcal{L}_{q2t}$   & $\mathcal{L}_{t2t}$ &   $\mathcal{L}_{p2p}$       &&&                 \\ \midrule
        &\checkmark   & &                                       &   83.76 &  79.52   & 81.64 \\
        \multirow{10}*{\shortstack{w/ \\ pretrain}} &\checkmark   &\checkmark &                                           &    82.87  &    81.43   & 82.15 \\
         &\checkmark   &  &  \checkmark &    83.74  &  80.27     & 82.00 \\
        
        & \checkmark  &\checkmark &\checkmark   &      84.0&     81.39&   82.70\\
        \midrule
        &\checkmark   &        &  &  85.12 &     81.2     &  83.16\\
        & \checkmark  & \checkmark   &    &85.52 &  81.92   &   83.72     \\
        &\checkmark   &       & \checkmark &  85.29      &   82.42   & 83.86  \\
        &\checkmark   &\checkmark   & \checkmark &85.55 & 82.71  & 84.13 \\
        \bottomrule
        
        \end{tabular}
        }   
        \label{tab:abmethod}
        }
        \hfill
        \subfloat[][Effectiveness of Pretraining]{
        \resizebox{0.48\linewidth}{!}{
            \begin{tabular}{l|c|cc|cc|c}
            \toprule
            {\multirow{2}{*}{Method}}  & \multirow{2}{*}{\begin{tabular}[c]{@{}c@{}}Pre-train with\\ ${\text{CIRR}_\text{MTST}}$ \end{tabular}}& \multicolumn{2}{c|}{$\text{Recall@K}$}      & \multicolumn{2}{c|}{$\text{Recall}_\text{subset}\text{@K}$}&\multirow{2}{*}{Avg. }  \\ \cmidrule{3-6} 
            {}                        &                                                                           & K=1      & K=5 & K=1         & K=2         \\ 
            \midrule
            {TIRG~\cite{tirg}}            & -   &          
            10.62	&38.36	&39.47&	61.05	&38.91\\
            {TIRG~\cite{tirg}}      & \checkmark   &          
            \textbf{18.60} & \textbf{53.54}  & \textbf{51.58}   & \textbf{72.61}& \textbf{52.56${\color{red}\uparrow 13.65}$}  \\
            {ARTEMIS~\cite{atermis}}            & -   &         
            17.47	&47.31&	40.70&	61.91	&44.00\\
            {ARTEMIS \cite{atermis}}      & \checkmark   &          
            \textbf{28.08} & \textbf{62.77}  & \textbf{53.75}   & \textbf{74.49}& \textbf{58.26${\color{red}\uparrow 14.26}$}  \\
            
            {CLIP4CIR \cite{clip4cir2}} & -   & 42.17 & 76.11  &69.70 & 87.42& 72.91 \\
            
            {CLIP4CIR \cite{clip4cir2}}      & \checkmark   &         	
            \textbf{44.69} & \textbf{77.57}  & \textbf{71.80}   & \textbf{88.14}& \textbf{74.68${\color{red}\uparrow 1.77}$}  \\
            {SPRC \cite{sprc}} & -  &                                                                  
            53.67	&82.87	&	81.44	&92.97	&	82.16\\
            {SPRC \cite{sprc}}     & \checkmark   &          
            \textbf{55.30} & \textbf{85.05}  & \textbf{81.46}   & \textbf{93.22}& \textbf{83.31${\color{red}\uparrow 1.15}$}  \\
            
            \bottomrule
            \end{tabular}
        }
        \label{tab:method}  
        }
    \end{minipage}
    \hfill
    \vspace{-0.2cm}
    \begin{minipage}[t]{1\textwidth}
        \subfloat[][{Comparison of the results using original text and generated text for triplets.}]{
        \resizebox{0.48\linewidth}{!}{
        \begin{tabular}{l|c|cc|cc|c}
        \toprule
        {\multirow{2}{*}{Method}}  & \multirow{2}{*}{Modifier}& \multicolumn{2}{c|}{$\text{Recall@K}$}      & \multicolumn{2}{c|}{$\text{Recall}_\text{subset}\text{@K}$}&\multirow{2}{*}{Avg. }  \\ \cmidrule{3-6} {}   &       & K=1      & K=5 & K=1         & K=2         \\ 
        \midrule
        {PTHA} & orgin  &56.80	&85.55	&	82.71	&94.00	&	84.13\\
        {PTHA}     & generated &
        \textbf{69.17} & \textbf{93.63}  & \textbf{88.14}   & \textbf{96.22}& \textbf{90.89}  \\
        \bottomrule
        \end{tabular}
        }
        \label{tab:val}  
        }
        \hfill
        \subfloat[][Comparable performance after pretraining]{
        \resizebox{0.48\linewidth}{!}{
            \begin{tabular}{cc|ccccccc|c}
            \toprule
             {CIRR}&  ${\text{CIRR}_\text{MTST}}$ & \multicolumn{4}{c|}{Recall@K}                                                                            & \multicolumn{3}{c|}{${\text{Recall}_{\text{subset}}\text{@K}}$}                                        & \multicolumn{1}{c}{Avg.} \\ \cmidrule{3-9}
              &  & \multicolumn{1}{c}{K=1} & \multicolumn{1}{c}{K=5} & \multicolumn{1}{c}{K=10} & \multicolumn{1}{c|}{K=50} & \multicolumn{1}{c}{K=1} & \multicolumn{1}{c}{K=2} & \multicolumn{1}{c|}{K=3} & \multicolumn{1}{c}{}                      \\ \midrule
            \checkmark &    &   53.03  &   83.76  &  90.60 &  97.96    &  79.52   &   92.71  &  96.82   & 81.64\\
             & \checkmark  & 51.28&  79.86& 87.75 &97.22&       77.01&    91.05&      96.17&   78.44\\
            \bottomrule
            \end{tabular}
            }
            \label{tab:zero}
        }
    \end{minipage}
    \hfill
    \vspace{-0.2cm}
    \begin{minipage}[t]{1\textwidth}
    \subfloat[][{Two stages' result comparison of PTHA and SPRC\cite{sprc}}]{
        \resizebox{0.48\linewidth}{!}{
            \begin{tabular}{l|c|cc|cc|c}
            \toprule
            {\multirow{2}{*}{Method}}  & \multirow{2}{*}{\begin{tabular}[c]{@{}c@{}}Query \\ \end{tabular}} & \multicolumn{2}{c|}{$\text{Recall@K}$}      & \multicolumn{2}{c|}{$\text{Recall}_\text{subset}\text{@K}$}&\multirow{2}{*}{Avg. }  \\ \cmidrule{3-6} 
            {}                        &                                                                           & K=1      & K=5 & K=1         & K=2         \\ 
            \midrule
            {SPRC \cite{sprc}} & implicit prototype ${f_{r2t}}$                                                                     &
            48.51&79.81	&	74.67	&90.17	&	77.24\\
            
            {PTHA}     &implicit prototype ${f_{r2t}}$                                                                     & 
             \textbf{51.21} & \textbf{82.09}  & \textbf{76.27}   & \textbf{90.74}& \textbf{79.18}  \\
            {SPRC \cite{sprc}}           & final feature ${f_q}$                                                                     & 
            56.32&85.50	&82.09	&93.49	&83.80\\
            
            {PTHA}      & final feature ${f_q}$                                                                      & 
            \textbf{56.80} & \textbf{85.55}  & \textbf{82.71}   & \textbf{94.00}& \textbf{84.13}  \\
            
            \bottomrule
            \end{tabular}
            }
            \label{tab:sprc}  
        }
        \hfill
         \subfloat[][{Zero-shot CIR performance on CIRCO\cite{circo} test set.}]{
        \resizebox{0.48\linewidth}{!}{

            \begin{tabular}{l|c|c|ccc}
            \toprule
            \multirow{2}{*}{Arch}&\multirow{2}{*}{Pretraining Method}&\multirow{2}{*}{Finetuning Method} &\multicolumn{3}{c}{$\text{mAP}\text{@K}$}\\
            \cmidrule{4-6} 
                    &&           & K=5      & K=10 & K=25              \\ 
            \midrule
            ViT-g/14 &CompoDiff\cite{compodiff}&-&15.33&17.71 & 19.45 \\
            
            ViT-g/14 &-&SPRC\cite{sprc}&19.68& 20.73 & 22.63 \\
            ViT-g/14 &${\text{CIRR}_\text{MTST}}$&PTHA&20.06&21.05&23.01\\
            ViT-g/14 &${\text{CIRR}_\text{MTST}}$&-& 22.8& 24.02 & 26.17\\
            {ViT-g/14} &{LinCIR}\cite{gu2024lincir}&-&20.34&21.85 & 23.98 \\
            {ViT-g/14} &{LDRE+IP-CIR}\cite{yang2024ldre}&-&32.75&34.26&36.86 \\
            {CoCa-L} &{MagicLens-L}\cite{magiclens}&-&\textbf{34.1}&\textbf{35.4} & \textbf{38.1} \\
            
            \bottomrule
            \end{tabular}

    }
        \label{tab:circo}
        }
    \end{minipage}
    
    \label{tab:sunmmary}
    \vspace{-0.6cm}
\end{table*}

\begin{figure}[!t]
    \vspace{-0.3cm}
    \centering
    \includegraphics[width=1\linewidth]{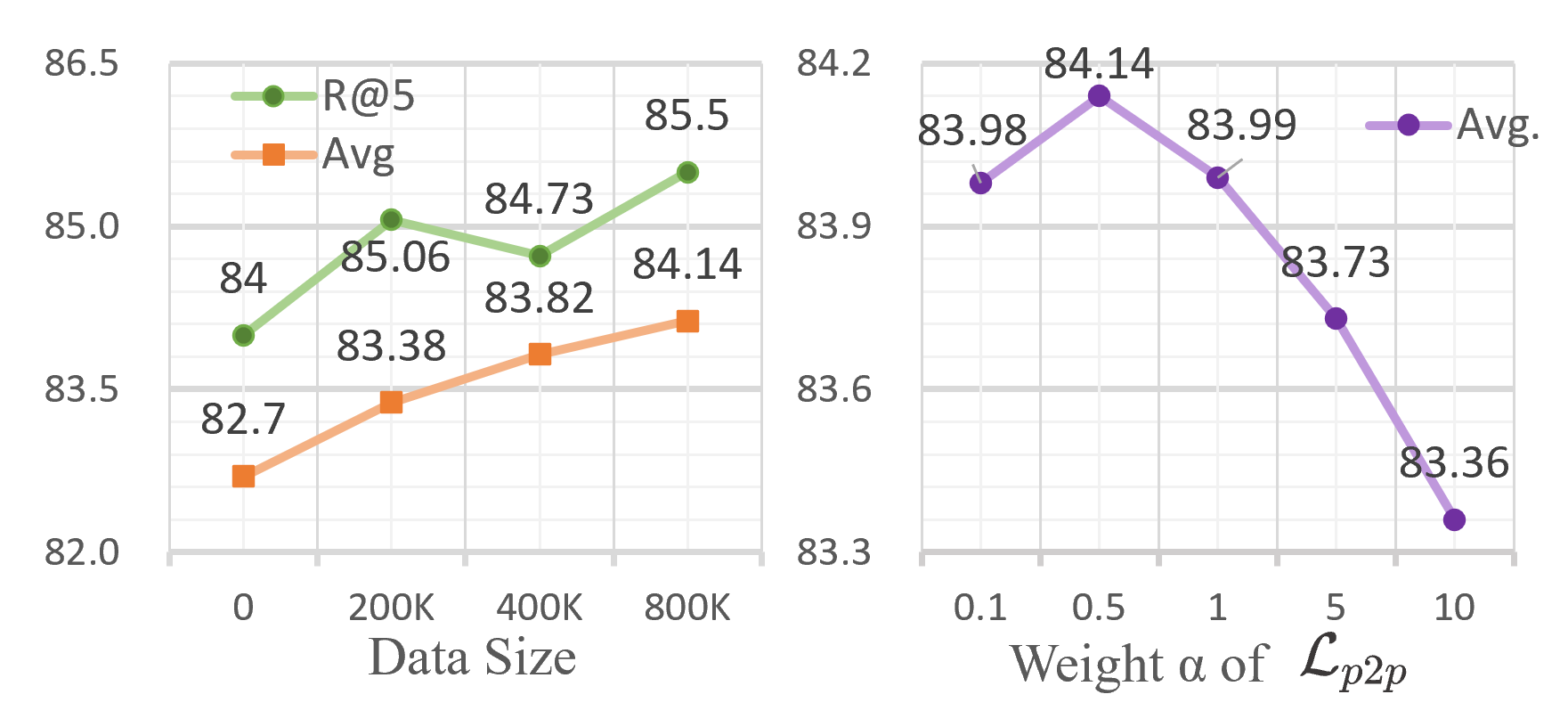}
    \vspace{-0.3cm}
    \caption{Left: Ablation studies on different pre-training ${\text{CIRR}_\text{MTST}}$ data size. We report Recall@5 and Avg. metric on CIRR validation set by fine-tuning with~\cref{finaloss}. Right: We deploy different ${L_{p2p}}$ weight ${\alpha}$ on fine-tuning stage with the same pre-trained model.}
    \label{zxt}
    \vspace{-0.4cm}
\end{figure}

\begin{figure}[!t]
    \centering
    \includegraphics[width=0.6\linewidth]{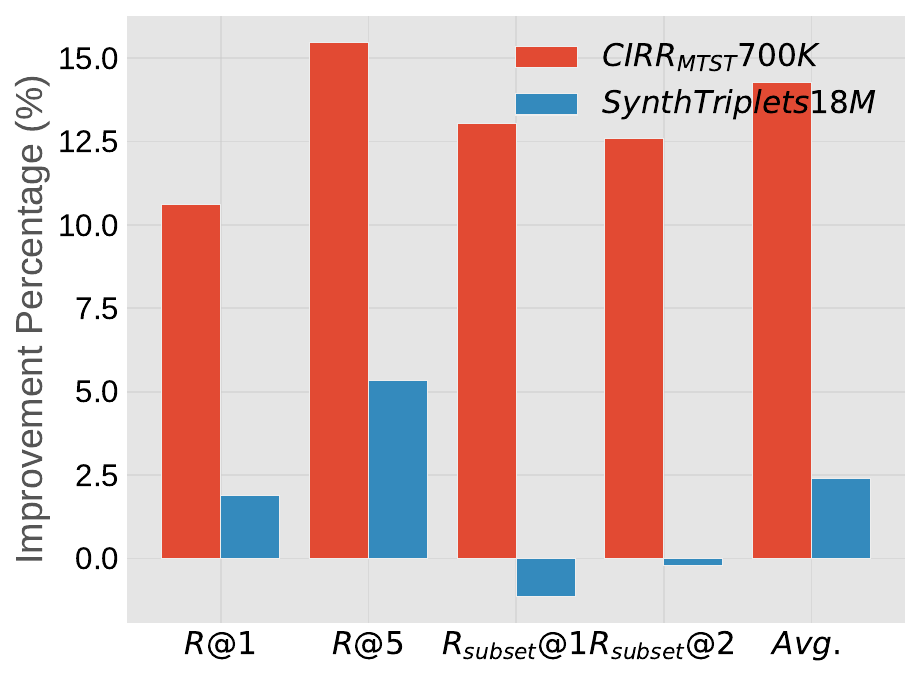}
    \caption{
    Two pre-training data, ${\text{CIRR}_\text{MTST}}$700K and SynthTriplets18M benefit on ARTEMIS {after same finetuning on CIRR}. With much less data, ${\text{CIRR}_\text{MTST}}$700K's performance enhancement on all markers is more comprehensive and significant than SynthTriplets18M's.}
    \label{fig:pruning_impact}
    \vspace{-0.6cm}
\end{figure}

\begin{figure*}[t]
  \centering
    \vspace{-0.2cm}
  \includegraphics[width=0.95\linewidth]{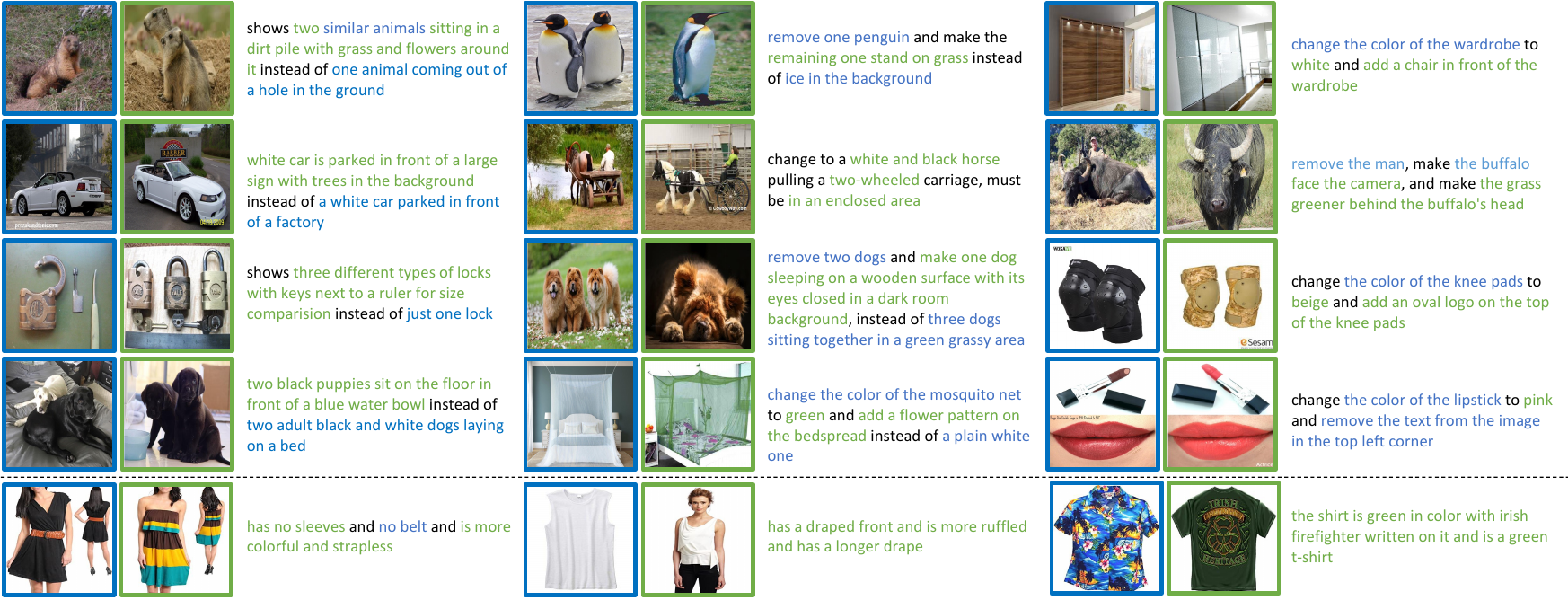}
  \caption{\textbf{Selected examples of generated triplets in ${\text{CIRR}_\text{MTST}}$(row 1-4) and ${\text{FashionIQ}_\text{MTST}}$(row5).} The \textcolor[HTML]{0070c0}{blue box} represents the reference image, while the \textcolor[HTML]{70ad47}{green box} indicates the target image. We leverage these two images as input to generate modified text. The \textcolor[HTML]{0070c0}{blue text} represents the information derived from the reference image, while the \textcolor[HTML]{70ad47}{green text} represents new additions or changes specific to the target image.}
    \vspace{-0.2cm}
  \label{fig:gensamples}
\end{figure*}

\begin{figure*}[t]
  \centering
    \vspace{-0.2cm}
  
  \includegraphics[width=0.95\linewidth]{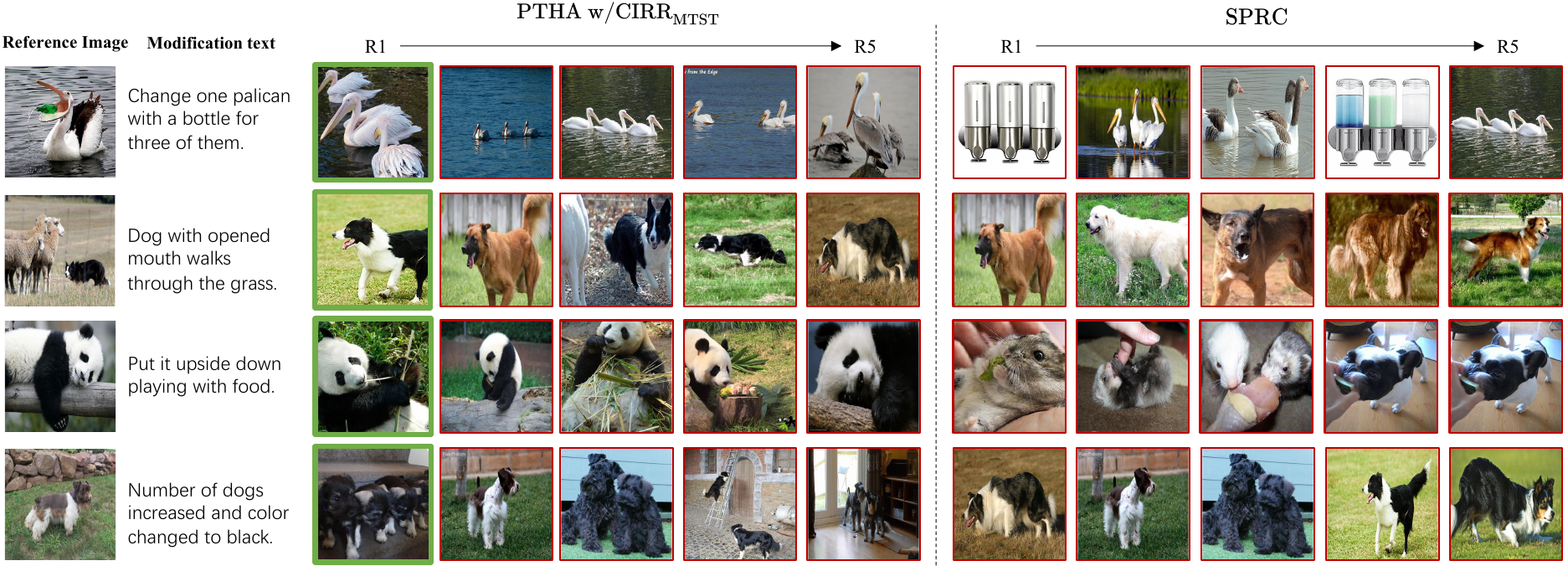}
  \caption{Qualitative CIR results of our methods and SPRC, placed in descending order from right to left based on similarity. The \textcolor[HTML]{70ad47}{green boxes} indicate the correct matches, and the images in the \textcolor[HTML]{aa0d06}{red boxes} are the wrong matches.}
  \label{fig:samples}
    \vspace{-0.2cm}
  
\end{figure*}

\subsection{Ablation Studies}

\subsubsection{Effect of PTHA Learning}
Table ~\ref{tab:abmethod} presents an analysis of the loss terms employed in PTHA learning, elucidating their contributions. Results from both the pretraining and finetuning phases confirm the effectiveness of all three losses. For instance, using only the query-to-target contrastive loss ($\mathcal{L}_{q2t}$) without pretraining yields an average performance of ${81.64}$. The inclusion of the text-to-target image term ($\mathcal{L}_{t2t}$) enhances this to ${82.15}$. The simultaneous deployment of all three losses optimizes performance to its peak. Under the cases of learning with pretraining, the contribution of each loss is sufficiently validated, and simultaneously applying three losses promotes our performance to the new state-of-the-art.  

\subsubsection{Effect of MTST pretraining}
By comparing the results of the cases with and without pretraining, it can be observed that regardless of the type of the applied loss function, the models that have undergone pretraining consistently exhibit significant performance improvement. 
In the application of four different combinations of loss functions, the pretraining respectively leads to an improvement of {1.52, 1.57, 1.86, and 1.43} in ``Avg.''. The effect of pretraining on enhancing Recall@K is notably substantial.
We also give a comprehensive discussion in terms of the size and the image source of pretraining MTST.  The results are reported in Figure \ref{zxt}, left.
As the size of the data increases from 0 to 800K, the performance of the model shows an upward trend.

\subsection{MTST Assessment}
\subsubsection{Scalability of MTST Pre-training}
To further validate the effectiveness of MTST Pre-training, we pre-train four more methods \cite{sprc,atermis,tirg,clip4cir2} using ${\text{CIRR}_\text{MTST}}$700K, which doesn’t contain the image pairs from the original image set.
The results are reported in Tab.\ref{tab:method}. MTST pre-training brings clear performance improvements across all four methods, especially TIRG\cite{tirg} and ARTEMIS\cite{atermis}.

\subsubsection{Generated Data Quality}

As shown in Table ~\ref{tab:zero}, by directly using our ${\text{CIRR}_\text{MTST}}$ for training without further fine-tuning, and excluding image pairs from the CIRR training set, we demonstrate competitive results directly on the validation set. This suggests a high similarity between ${\text{CIRR}_\text{MTST}}$ and those of CIRR. 
To further assess the quality of generated data, we generate 4K modifiers from the CIRR validation image pairs. Our evaluation includes 3 parts: 

(1) Direct evaluation. The generated modifier text can achieve a 0.25 ROUGE-1 Score and 0.19 METEOR Score. However, due to the modifiers' diversity, we affirm these metrics don't fully measure the quality. 

(2) Indirect Evaluation.
Substituting generated text for CIRR validation leads to much higher scores, highlighting our data's robust quality and pre-training suitability (see Table ~\ref{tab:val}).

(3) User study. We ask five experts to assess 100 randomly selected image pairs in ${\text{CIRR}_\text{MTST}}$ to choose the better text between the real modifier and the generated modifier. We randomly shuffle the order of these two types of text. The findings suggest a comparable preference for generated text over real text, with favorable scores of 43\%, 40\%, 50\%, 46\%, and 45\%, respectively. 
It is worth noting that we provide users with an option to report any factual errors in the two modification texts corresponding to each image pair, and we report the average error reporting rate, which is calculated as: the number of reported errors in the generated modification texts /(the total number of generated modification texts ${\times }$the number of test users). The final average error reporting rate of generated text is only 4\%. 

\subsubsection{Comparison with other generated data}
We compare the gains of the ${\text{CIRR}_\text{MTST}}$700K vs. SynthTriplets18M \cite{compodiff} on ARTEMIS\cite{atermis}.{ We separately utilize these two data to pretrain and CIRR training set to finetune.}
Figure \ref{fig:pruning_impact} shows that with much less data, our performance enhancement on all markers is more comprehensive and significant than SynthTriplets18M.

\subsection{PTHA Assessment}
\noindent Our method shares the same baseline with \cite{sprc}.  
We observe PTHA's comparable performance in Table ~\ref{tab:cirrresult}, row 16.
Notably, our PTHA, combined with pre-training, shows more than additive effectiveness, especially on Recall@1.
When fine-tuned with an identical pre-trained model, PTHA outperforms SPRC on CIRR (See Table ~\ref{tab:sprc}, rows 3-4). This is due to the consistency of the generated data in the pre-training and fine-tuning stages, as well as the supervisory role of the $\mathcal{L}_{p2p}$.
Further more, PTHA has a better intermediate feature quality. We directly utilize the features extracted solely through the multi-modal encoder (i.e., the {\bf implicit prototype}, ${f_{r2t}}$.) as the query for validation. The results, as shown in Table ~\ref{tab:sprc}, rows 1-2, clearly indicate that our first-phase feature quality is superior to SPRC's.

{We set the weight of ${\mathcal{L}_{q2t}}$ to 1 and the weight of ${\mathcal{L}_{t2t}}$ to 0.4 following SPRC\cite{sprc}. We conduct experiment with different weight $\alpha$ of ${\mathcal{L}_{p2p}}$(See Figure \ref{zxt}, right) on pre-trained model using ${\text{CIRR}_\text{MTST}}$700k. Weighting ${\mathcal{L}_{p2p}}$ with ${\alpha=0.5}$ leads to the best results.}

{\subsection{Zero-shot ability on CIRCO\cite{circo}}}
{\noindent As shown in Table ~\ref{tab:circo}, we compare the performance of our pretrained model, fine-tuned model, SPRC\cite{sprc}, {and other pretrained models\cite{compodiff,magiclens,gu2024lincir,yang2024ldre}} on CIRCO\cite{circo} in a zero-shot setting. It can be seen that our fine-tuned model outperforms SPRC, and our pre-trained model shows stronger generalization capabilities with higher zero-shot performance.}

\subsection{Qualitative Results}
\noindent As shown in Figure \ref{fig:gensamples}, we qualitatively present the triplets generated in ${\text{CIRR}_\text{MTST}}$ and ${\text{FashionIQ}_\text{MTST}}$. Particularly in ${\text{CIRR}_\text{MTST}}$, our generated modification text encompasses both a description of the target image and the changes observed by comparing the two images. We have observed that the model demonstrates strong descriptive capabilities in capturing changes in features such as color, condition, quantity, and the addition or removal of objects. Furthermore, we also compare the retrieval results of our method with the state-of-the-art approach, SPRC\cite{sprc} on several examples (see Figure \ref{fig:samples}). We can observe that during the retrieval process, our method effectively preserves the relevant implicit prototypes of the reference image based on the modification text. The target images largely retain these implicit prototypes. For instance, in the first row, when the description might mislead the model, we accurately preserve the ``pelican'' prototype. In the second to fourth rows, we implicitly retain the characteristics of animals from the reference image.

\section{Conclusion}
\noindent In this paper, we focus on alleviating the scarcity of training triplets in composed image retrieval. To this end, we train a modification text generator that produces synthetic, high-quality modification-oriented triplets.  Our generator inputs two images and outputs versatile, descriptive modifications to form realistic-like triples. With the trained generator, we benefit from the learning of CIR in both the pretraining and fine-tuning stages. In the pretraining stage, we generate the large-scale triplets to perform pretraining. In the fine-tuning stage, we first synthesize the reversed modification text, supporting us design a two-step alignment mechanism to gradually address the gap between the multimodal query and the target image. We first learn the implicit prototype with the real triplet and its reverse counterpart and combine the implicit prototype with the modification text to align with the target image. Extensive experimentation on benchmark datasets from both natural and fashion domains demonstrates that our method achieves a comparable performance with state-of-the-art approaches.

\section*{Limitations And Future Work}
{
\noindent Our work primarily offers a paradigm for expanding training sets on CIR. Focusing on improving performance on specific datasets, CIRR\cite{cirr} and FashionIQ\cite{fashioniq}, the model is pre-trained on the extended data of such dataset. Therefore, training on specific domain results in insufficient effectiveness and generalization compared to previous methods that utilize large-scale triplet pre-training, as shown in zero-shot CIR performance on CIRCO\cite{circo} in Table \ref{tab:circo}. Furthermore, applying MTST generation strategy to MLLMs\cite{Qwen2VL,chen2024internvl} that already possess the ability to distinguish between two images leads to a decrease in the model's generalization capability. Future work could benefit from employing powerful MLLMs to further explore a more rapid and efficient approach to domain adaptation, along with a method for generating text with higher quality and finer-grained modifications.}

\bibliographystyle{IEEEtran}
\bibliography{egbib}

\begin{IEEEbiography}
[{\includegraphics[width=1in,height=1.25in,clip,keepaspectratio]{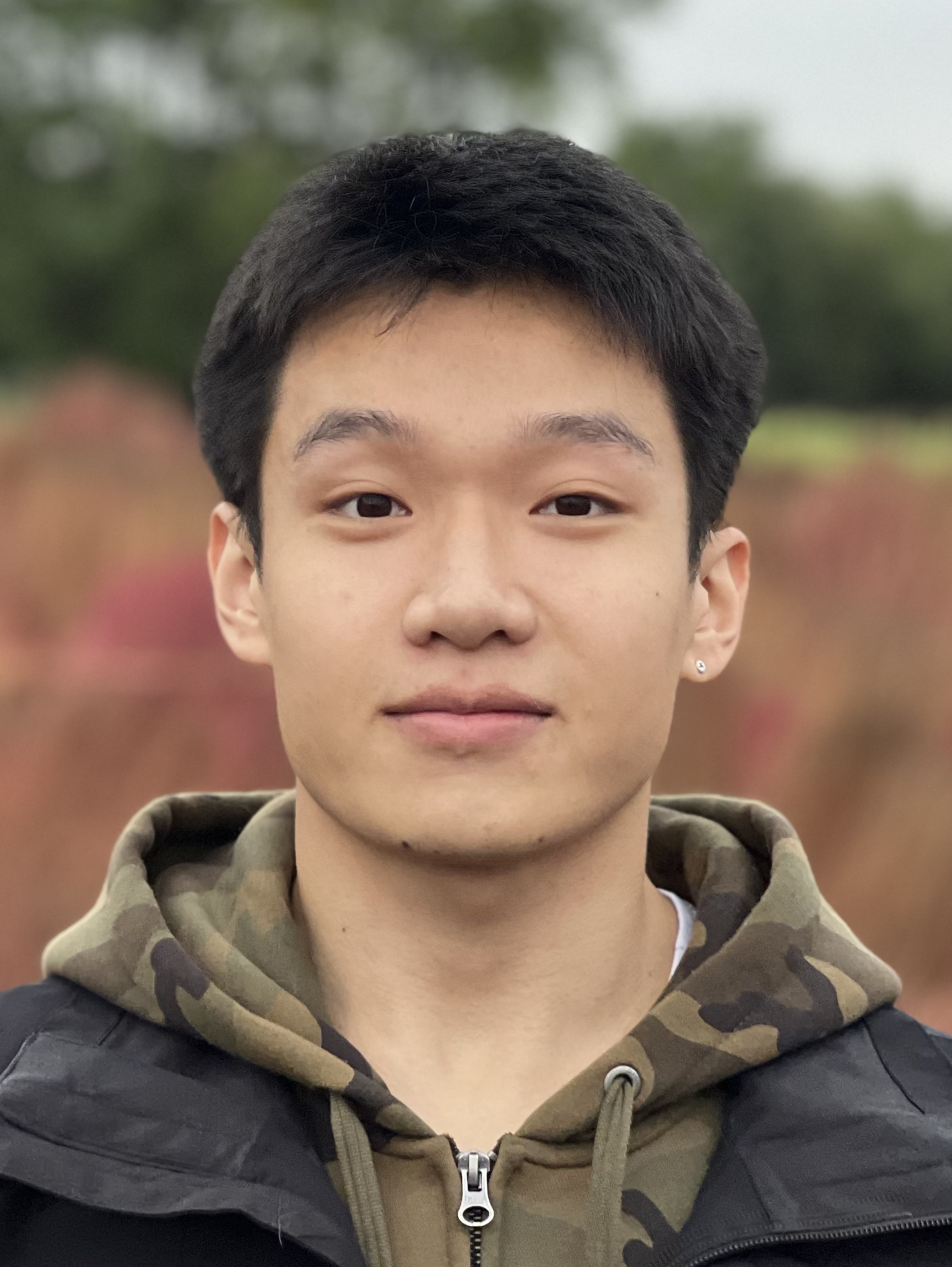}}]{Yinan Zhou}received the B.S. degree from School of software Engineering, Xi’an Jiaotong University, Xi’an, China, in 2021. He is now a Ph.D. student in the School of Software Engineering at Xi'an Jiaotong University. His research interests include composed image retrieval, cross-modal retrieval, and multi-modal large language model.
\end{IEEEbiography}
\vspace{-1.08cm}

\begin{IEEEbiography}[{\includegraphics[width=1in,height=1.25in,clip,keepaspectratio]{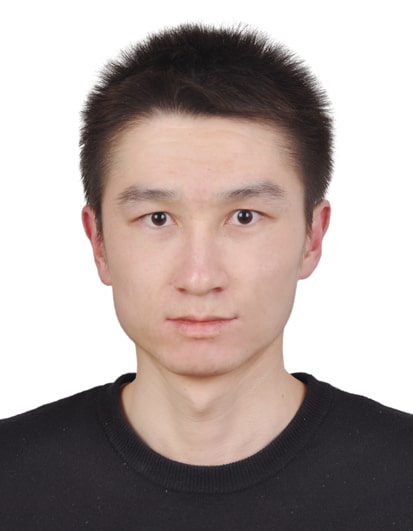}}]{Yaxiong Wang} received the B.S. degree from Lanzhou University, Lanzhou, China, in 2015, and Ph.D. degree at School of software Engineering, Xi’an Jiaotong University, Xi’an, China, in 2021. He is now a associate professor in Hefei University of Technology. His research interests include cross-modal retrieval, image generation, semantic segmentation, and ReID.
\end{IEEEbiography}
\vspace{-1.08cm}

\begin{IEEEbiography}[{\includegraphics[width=1in,height=1.25in,clip,keepaspectratio]{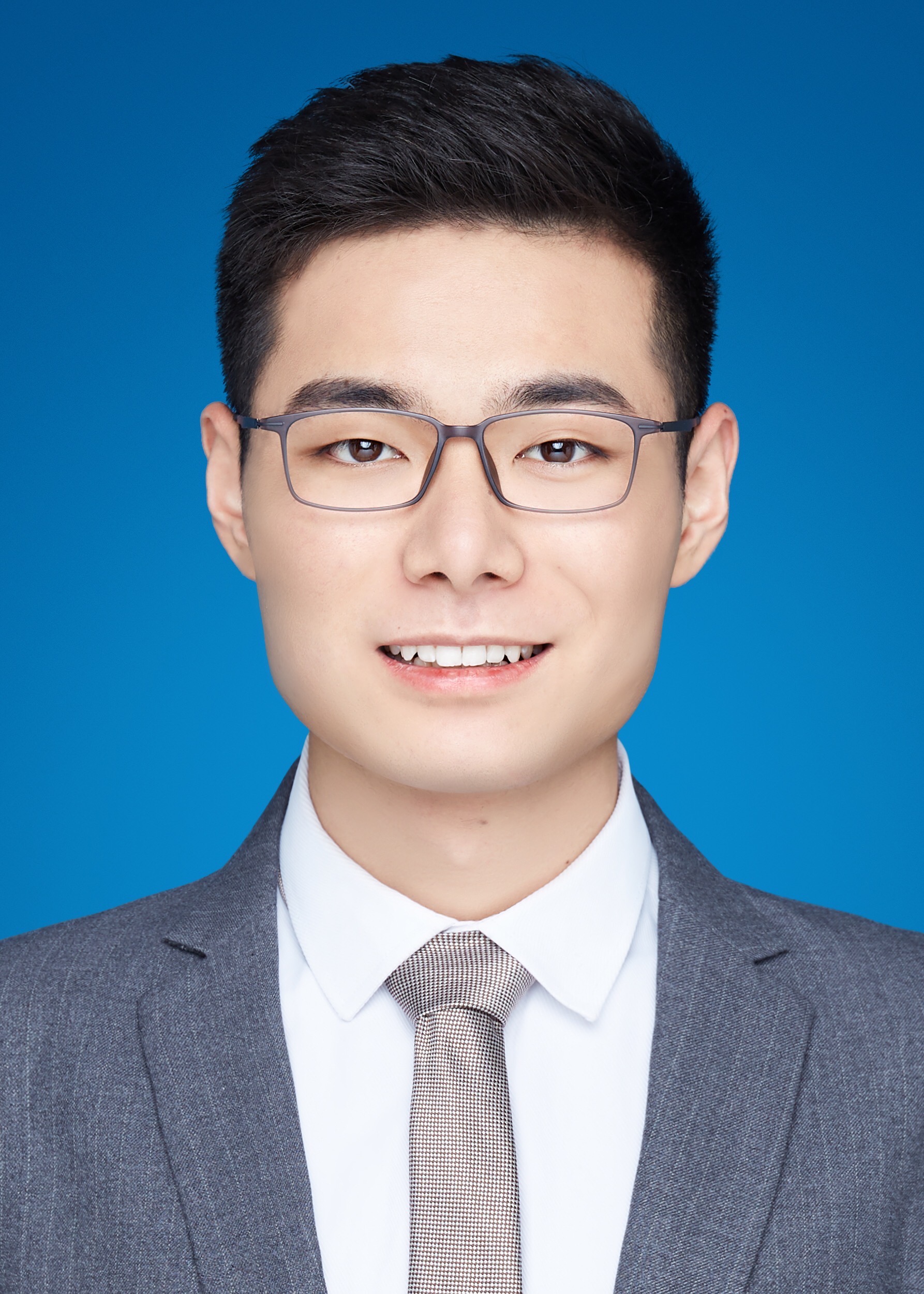}}]{Haokun Lin} received the B.S. degree from School of Software Engineering, Huazhong University of Science and Technology, Wuhan, China, in 2021. He is now a Ph.D. student in Institute of Automation, Chinese Academy of Sciences. His research interests include multi-modal learning, model compression and large language models.
\end{IEEEbiography}
\vspace{-1.08cm}

\begin{IEEEbiography}
[{\includegraphics[width=1in,height=1.25in,clip,keepaspectratio]{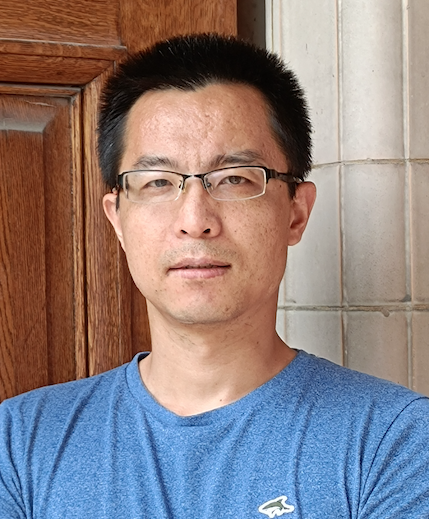}}]{Chen Ma} got his PhD from the School of Computer Science, McGill University, supervised by Prof. Xue (Steve) Liu. In the meantime, he also closely worked with Prof. Mark Coates. Before joining McGill, he received his MS and BS degrees in Software Engineering from Beijing Institute of Technology.
He is currently an Assistant Professor in the Department of Computer Science, at the City University of Hong Kong since August 2021. His main research interests include natural language processing, recommender systems and data mining.
\end{IEEEbiography}
\vspace{-1.08cm}

\begin{IEEEbiography}
[{\includegraphics[width=1in,height=1.25in,clip,keepaspectratio]{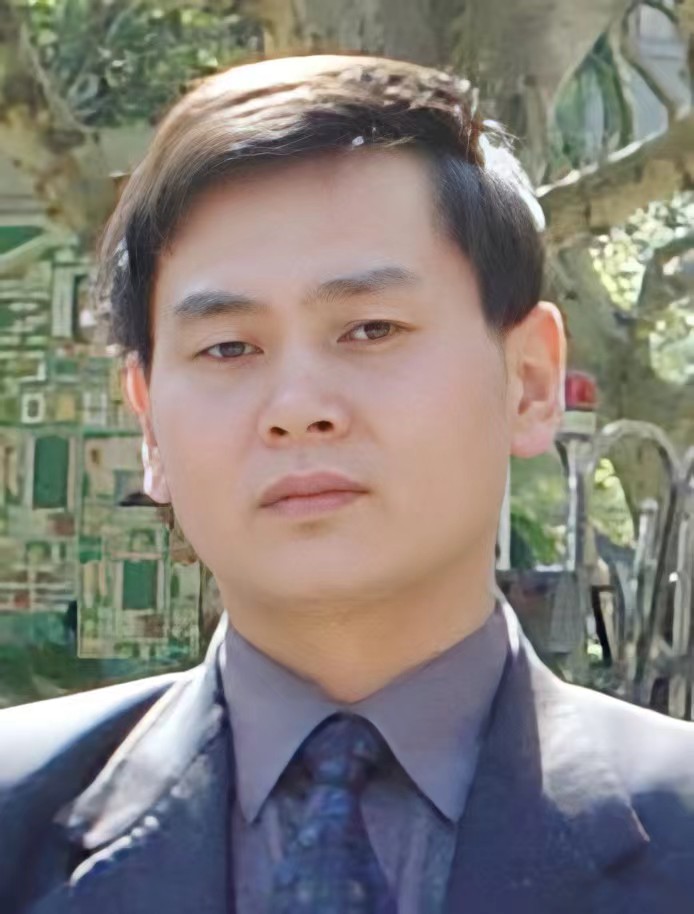}}]{Li Zhu}
received the B.S. degree from Northwestern Polytechnical University, Xi’an, China, in 1989, and the M.S. and Ph.D. degrees from Xi’an Jiaotong University, Xi’an, in 1995 and 2000, respectively.,He is currently a Professor with the School of Software, Xi’an Jiaotong University. His main research interests include multimedia processing and communication, parallel computing, and networking.
\end{IEEEbiography}
\vspace{-1.08cm}

\begin{IEEEbiography}[{\includegraphics[width=1in,height=1.25in,clip,keepaspectratio]{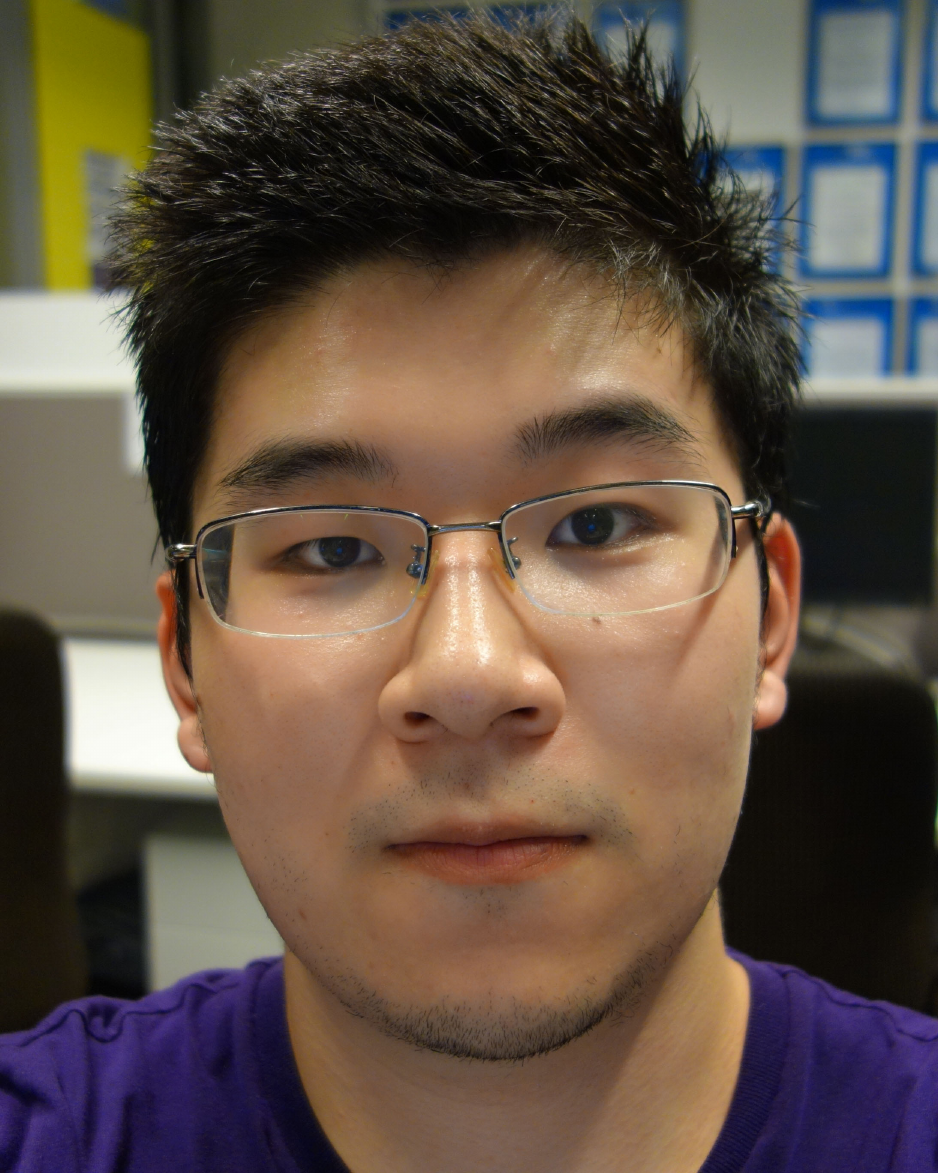}}]{Zhedong Zheng}
is an Assistant Professor with the University of Macau. 
He received the Ph.D. degree from the University of Technology Sydney in 2021 and the B.S. degree from Fudan University in 2016. He was a postdoctoral research fellow at the School of Computing, National University of Singapore. He received the IEEE Circuits and Systems Society Outstanding Young Author Award of 2021. 
His research interests include robust learning for image retrieval, generative learning for data augmentation, and unsupervised domain adaptation. He served as the senior PC for IJCAI and AAAI, and the area chair for ACM MM'24 and ICASSP'25.
\end{IEEEbiography}
\vspace{-1.08cm}

\vfill

\end{document}